\documentclass[11pt,english,authoryear,review]{elsarticle}
\usepackage{lmodern}
\usepackage{lmodern}
\usepackage[T1]{fontenc}
\usepackage[latin9]{inputenc}
\usepackage{geometry}
\usepackage{amsmath}
\usepackage{amssymb}
\usepackage{graphicx}
\usepackage{setspace}
\usepackage{epstopdf}
\makeatletter

\providecommand{\tabularnewline}{\\}

\journal{arXiv}
\usepackage{tikz}
\usepackage{ulem}\bibpunct{(}{)}{,}{a}{,}{,}

\usepackage{babel}
\usepackage{tikz}

\usepackage{babel}

\@ifundefined{showcaptionsetup}{}{%
 \PassOptionsToPackage{caption=false}{subfig}}
\usepackage{subfig}
\makeatother

\usepackage{babel}
\begin{document}

\begin{frontmatter}{}

\title{On the spatial convergence and transient behaviour of lattice Boltzmann
methods for modelling fluids with yield stress}

\author[uq,mel]{W.~Regulski\corref{cor1}}

\ead{wregulski@meil.pw.edu.pl}

\author[uq]{C. R. Leonardi}

\author[mel]{J. Szumbarski}

\cortext[cor1]{Corresponding author}

\address[uq]{School of Mechanical and Mining Engineering. The University of Queensland, Australia}

\address[mel]{Institute of Aeronautics and Applied Mechanics. Warsaw University of Technology, Poland}

\begin{abstract}
In this paper, the performance of two lattice Boltzmann method formulations for yield-stress (i.e. viscoplastic) fluids has been investigated. The first approach is based on the popular Papanastasiou regularisation of the fluid
rheology in conjunction with explicit modification of the lattice Boltzmann relaxation rate. The second approach uses a locally-implicit formulation to simultaneously solve for the fluid stress and the underlying particle distribution functions. After investigating issues related to the lattice symmetry and non-hydrodynamic Burnett stresses, the two models were compared in terms of spatial convergence and their behaviour in transient and inertial flows. The choice of lattice and the presence of Burnett stresses was found to influence the results of both models, however the latter did not significantly degrade the velocity field. Using Bingham flows in ducts and synthetic porous media, it was found that the implicitly-regularised model was superior in capturing transient and inertial fluid behaviour. This result presents potential implications for the application of the Papanastasiou-regularised model in such scenarios. In creeping flows the performance of both models was found to be both similar and satisfactory. 
\end{abstract}


\end{frontmatter}{}

\section{Introduction}
\label{sec:Introduction}

Yield-stress fluids, also known as viscoplastic fluids, are characterised by the property that they remain unyielded below a specific value of yield stress, $\sigma_{y}$. This type of behaviour is observed in many common (e.g. ketchup, paints, toothpaste) and engineering (e.g. crude oils, concrete) materials, as well as in nature (e.g. blood, muds). This property can also be exploited in the design of bespoke fluids for suspension transport in porous and fractured media, such as those used in the production of oil and gas. The prevalence of viscoplastic fluids in science and engineering necessitates the development of robust numerical tools that can be used to predict and optimise their behaviour in arbitrary flow configurations.

Although the existence of an actual value of yield stress has been the subject of discussion \citep{Barnes1999_panta_rei}, the idea of using a specific yield stress value in constitutive modelling has been extensively utilised. The simplest model of yield-stress liquids is the Bingham fluid \citep{Bingham_1916}, which is characterised by its yield stress and plastic viscosity, $\mu_{p}$. The rheological relation takes the form,

\begin{equation}
\begin{cases}
\begin{array}{c}
\boldsymbol{\sigma}=2\left(\mu_{p}+\frac{\sigma_{y}}{\dot{\gamma}}\right)\boldsymbol{D},\\
\boldsymbol{D}=0,
\end{array} & \begin{array}{c}
\left|II_{\sigma}\right|\geq\sigma_{y}\\
\:\left|II_{\sigma}\right|<\sigma_{y}
\end{array}\end{cases},
\label{eq:rheoViscoplasticBase}
\end{equation}

\noindent in which $\mathbf{\boldsymbol{\sigma}}$ is the shear stress tensor, $\boldsymbol{D}$ is the rate of strain tensor,

\begin{equation}
\boldsymbol{D}=\frac{1}{2}\left(\nabla\boldsymbol{u}+\left(\nabla\boldsymbol{u}\right)^{T}\right),
\label{eq:rateOfStrain}
\end{equation}

\noindent and $II_{\sigma}$ is the second principal invariant of the shear stress tensor,

\begin{equation}
II_{\mathbf{\boldsymbol{\sigma}}}=-\frac{1}{2}\boldsymbol{\mathbf{\boldsymbol{\sigma}}}:\boldsymbol{\mathbf{\boldsymbol{\sigma}}}.
\label{eq:stressInvariant}
\end{equation}

\noindent The characteristic rate of strain, $\dot{\gamma}$, is calculated from the invariant of the rate of strain tensor, $II_{D}$, as,

\begin{equation}
\dot{\gamma}=2\sqrt{\left|II_{D}\right|}=\sqrt{2\boldsymbol{D}:\boldsymbol{D}}.
\label{eq:strainInvariant}
\end{equation}

\noindent This type of rheological relation belongs to the family of generalised Newtonian fluids with the apparent viscosity defined as a function of the characteristic rate of strain,

\begin{equation}
\mu_{app}=\mu_{p}+\frac{\sigma_{y}}{\dot{\gamma}}.
\label{eq:muApparent}
\end{equation}

Inspection of the Bingham model reveals two numerical difficulties. First, the model experiences an infinite apparent viscosity as the strain rate approaches zero. Secondly, it requires a distinction between the yielded and unyielded zone, where in the latter the medium must be treated as rigid. Thus, \textit{conventional} approaches to computational fluid dynamics such as the finite element method (FEM) face considerable difficulties in modelling this behaviour which are only overcome by the use of complex algorithms for mesh refinement and adaptivity \citep{Roquet_Saramito_2001_aFEM_pipe,Roquet_Saramito_2003_aFEM_cylinder}.
This can make such methods impractical from an engineering standpoint, particularly when simulating flows in which the topology of the structural boundaries is evolving.

\begin{figure}
\hfill{}\input{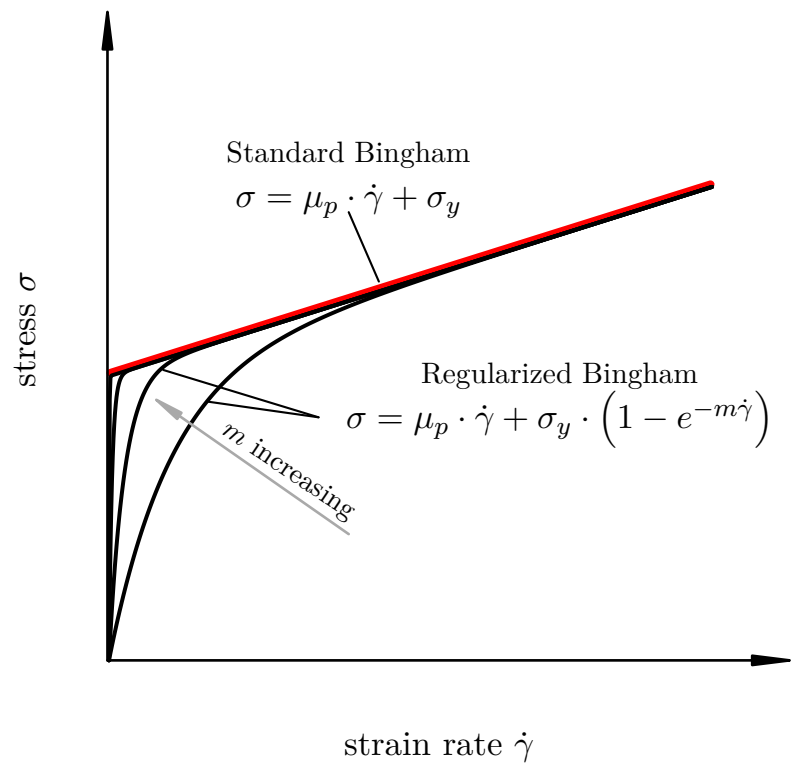}\hfill{}
\caption{Schematic plot of stress-strain rate dependence, $\sigma\left(\dot{\gamma}\right)$, for the standard Bingham model and the Papanastasiou-regularised (PR) model. A number of PR curves are plotted for values of $m=5,\,10,\,100,\,1000$ showing that as the regularisation parameter increases, the stress-strain curve tends to the line representing the standard Bingham fluid.}
\label{fig:BinghamSchematic}
\end{figure}

A common solution to the issues surrounding numerical implementation of Bingham fluids  is the regularisation procedure \citep{Papanastasiou1987}. The modified stress-strain relation then takes the form,

\begin{equation}
\boldsymbol{\sigma}=2\left(\mu_{p}+\frac{\sigma_{y}}{\dot{\gamma}}\left(1-e^{-m\dot{\gamma}}\right)\right)\boldsymbol{D}
\label{eq:rheoViscoplasticReg}
\end{equation}

\noindent where $m$ is the regularisation parameter which can attain very large values (e.g. $10^{5}-10^{10}$). However, $m$ must be $\mathcal{O}(10^{6})$ or greater to adequately produce viscoplastic behaviour (see Section \ref{subsec:Pap-reg-nu-dep}). This new rheological relation, as shown in Figure \ref{fig:BinghamSchematic}, relaxes the condition for the apparent viscosity, $\mu_{app}$, in the zero-strain limit so that it reaches a very large value instead of an infinite one,

\begin{equation}
\lim\mu_{app}\left(\dot{\gamma}\right)_{\dot{\gamma}\to0}=\mu_{p}+m\sigma_{y}
\label{eq:muApparentReg}
\end{equation}
 
\noindent Additionally, regularisation results in no unyielded zone in the flow. This approach, although giving reasonable results, has been criticised for its inability to produce the correct zero-strain zones \citep{Vikhansky2008_short_paper}. It has also been shown to produce inappropriate results in terms of the shape and topology of the unyielded zones, as highlighted by \cite{Roquet_Saramito_2001_aFEM_pipe}.

In spite of its limitations, the regularisation procedure, henceforth referred to as \textit{Papanastasiou-regularisation} (PR), has received much attention in the literature. Simple duct flows \citep{TaylorWilsonBicr112Reg,Guo_et_all_2011_MRT_nonnewtonian}, flows past arrays of obstacles \citep{Spelt2005}, as well as flows with particles \citep{China_2015_LBM_DEM} or with free surfaces \citep{Ginzburg2002_Bingham_freesurface} have been successfully simulated by means of classical numerical methods or alternative approaches such as the lattice Boltzmann method (LBM). The LBM has been especially successful as a consequence of some of its favourable properties such as straightforward implementation of complex geometrical boundaries and the ability to determine the rate of strain tensor locally \citep{Chen_Doolen_98_basicLBM}. It should be also noted that, instead of regularisation, some authors have used the LBM to model Bingham fluids simply by prescribing an upper bound for the fluid viscosity \citep{Svec2012Denmark,ALeonardi_2014_LBM_DEM_Binghamraey}.

The numerical construction of the LBM admits another approach to regularisation \citep{Vikhansky2008_short_paper}. Rather than modifying the fluid viscosity as in the PR model, the LBM collision process is constructed in a manner that permits the implicit calculation of fluid stress and the underlying particle distribution functions. This technique makes it possible to (nearly) impose the zero-strain condition or unmodified viscoplastic apparent viscosity by appropriate construction
of the second moment of the collision operator. In this study this model is referred to as \textit{implicitly-regularised} (IR).

This study presents a comparison of the PR and IR approaches to modelling yield-stress fluids in the context of the LBM. Particular attention is given to issues which are often overlooked or poorly understood such as convergence, transient and inertial behaviour, and some aspects of their numerical implementation. This work is divided into four main parts. Following this introduction, Section \ref{sec:NavierStokesLBM} reviews some aspects of the LBM that are important to this work and describes how it can be constructed in an implicit manner. Section \ref{sec:GenNewtonianLBM} then describes how the standard and implicit LBM constructions can be modified to simulate yield-stress fluids, with a focus on the Bingham model. Extensive results comparing the PR and IR models in duct flows and flows past a periodic array of circular cylinders are then presented in Section \ref{sec:Results}. The transient behaviour of both models is investigated and discussed, and some numerical aspects of the PR model are evaluated. In the context of unidirectional flows, the inaccuracy of the accepted analytical solution for the square duct is demonstrated, while some undesirable issues that originate from the kinetic nature of the LBM are also discussed. Finally, Section \ref{sec:Conclusion} presents some concluding remarks including avenues for future research.

\section{The lattice Boltzmann method for Navier-Stokes flows}
\label{sec:NavierStokesLBM}

The lattice Boltzmann equation (LBE) is a Boltzmann-type transport equation that governs the evolution of a certain set of discrete probability density functions, $f_{i}\left(t,\boldsymbol{x}\right),\;i=0,...,N-1$, which are often referred to as populations \citep{Chen_Doolen_98_basicLBM}. The LBM evolved from lattice gas cellular automata (LGCA), which was first formulated to recover the behaviour of gases \citep{Higuera_Jimenez_1989}, but it has subsequently been shown that the LBM is much more suited to the simulation of incompressible hydrodynamics. Currently, the LBM is viewed as a specific discretisation of the Boltzmann equation \citep{LSL_1997_BEtoLBE}. What differentiates the LBM from other discretisations of the Boltzmann equation is the  assumption that the particle populations travel via a specially chosen set of
velocities, $\boldsymbol{c}_{i}$. The LBE is then a partial differential equation system of the form,

\begin{equation}
\frac{\partial f_{i}}{\partial t}+\boldsymbol{c}_{i}\cdot\frac{\partial f_{i}}{\partial\boldsymbol{x}}=\Omega\left(\boldsymbol{f}\right),\;i=0,...,N-1,
\label{eq:8-1}
\end{equation}

\noindent in which $\Omega\left(\boldsymbol{f}\right)$ is the collision operator. The velocity set is constructed to provide this dynamic system with enough degrees of freedom to recover the governing equations of hydrodynamics. More specifically, the LBE is equivalent to the system of transport equations for density, $\rho$, velocity, $\boldsymbol{u}$, momentum flux, $\boldsymbol{\Pi}$, and higher order moments of particle distribution functions \citep{Dellar_2003_MRT_convergence}. The hydrodynamics is recovered at the macroscale with its basic quantities of density, velocity, and momentum flux emerging from the moments of the distribution functions,

\begin{subequations}
\begin{align}
\rho=\sum_{i=0}^{N-1}f_{i},
\end{align} 
\begin{align} 
\boldsymbol{u}=\frac{1}{\rho}\sum_{i=0}^{N-1}\boldsymbol{c}_{i}f_{i},
\end{align}
\begin{align} 
\mathbf{\mathbf{\boldsymbol{\Pi}}}=\sum_{i=0}^{N-1}\boldsymbol{c}_{i}\boldsymbol{c}_{i}f_{i}.  
\end{align}
\label{eq:MomentsLBM}
\end{subequations}

There is no direct incompressibility constraint in the LBE and fluid pressure is recovered from the equation of state, $p=\rho c_{s}^{2}$, with the lattice speed of sound, $c_{s}$, which is defined as $c_{s}=1/\sqrt{3}$ in most lattice constructions. On one hand, the compressible nature of the system is a very important advantage of this approach as there is no need to solve the pressure Poisson equation. On the other, it imposes a significant constraint which is that the incompressible dynamics is recovered in the limit of vanishing Mach number, $Ma=\left|\boldsymbol{u}\right|/c_{s}\rightarrow0$. It must be also noted that the hydrodynamic behaviour is only the slowly-varying solution of the LBE with respect to a certain time-scale associated with the collision operator \citep{Dellar_2003_MRT_convergence}. There exists dynamics of the higher-order terms associated with the fast-evolving part of the system. Its presence, which is very often neglected, manifests itself in some spurious (from the point of view of hydrodynamics) phenomena such as Burnett stresses \citep{DellarReis_2016_BurnetStress} or influences the computational stability of the method \citep{Geier2015_cumulant_LBM}. The presence of the Burnett stress turns out to have a significant impact on the viscoplastic models used in the LBM framework and this issue is discussed in detail in Section \ref{sub:Burnett_theory}.

The LBM operates on a certain spatial-temporal discretised form the the LBE \citep{Dellar2013_StrangSplitting},

\begin{equation}
f_{i}(t+\Delta t,\boldsymbol{x}_{j}+\boldsymbol{c}_{i}\Delta t)-f_{i}(t,\boldsymbol{x}_{j})=\Omega_{i}\left(\boldsymbol{f}\right),
\label{eq:8}
\end{equation}

\noindent in which $\boldsymbol{x}_{j}$ represents the points on the Cartesian grid and the time step, $\Delta t$, is chosen in a way that the advection part of the LBE is exact (i.e. particle populations travel from one grid point to another). An important point, very often not given sufficient attention, is that this discretised form of the LBE operates on a transformed set of populations,

\begin{equation}
\bar{f_{i}}=f_{i}-\frac{\Delta t}{2}\Omega_{i}.
\label{eq:pop_transform}
\end{equation}

\noindent This transformation renders the scheme implicit, although analytically solvable, and this property is extensively exploited during description of both viscoplastic models. Henceforth, however, the overbar will be excluded for brevity and the populations should be considered to be the transformed set, unless otherwise noted.

\subsection{Collision operators in the LBM}

A range of collision operators for the LBE have been proposed
\citep{matrix_reloded_2011,MRT2002,Cascaded_original_2006,Geier2015_cumulant_LBM,ELB_turbulence_2010}.
Most of them rely on the idea of relaxing the population values towards their equilibrium value, $f_{i}^{eq}$, usually expressed as the second-order truncation of the Maxwell-Boltzmann form,

\begin{equation}
f_{i}^{eq}=\rho w_{i}\left(1+\frac{\boldsymbol{u}\cdot\boldsymbol{c}_{i}}{c_{s}^{2}}+\frac{(\boldsymbol{u}\cdot\boldsymbol{c}_{i})^{2}}{2c_{s}^{4}}-\frac{\boldsymbol{u}^{2}}{2c_{s}^{2}}\right),
\label{eq:f_eq}
\end{equation}

\noindent with $w_{i}$ being the weights associated with specific lattice velocity vectors. Equilibrium distributions can also be derived from the assumption of the minimisation of certain functionals with constraints of mass and momentum conservation \citep{ELB_turbulence_2010,Geier2015_cumulant_LBM}. An important property of the equilibrium distribution is that its consecutive moments respectively yield mass, momentum and the momentum flux tensor,

\begin{subequations}
\begin{align}
\rho=\sum_{i=0}^{N-1}f_{i}^{eq},
\end{align} 
\begin{align} 
\boldsymbol{u}=\frac{1}{\rho}\sum_{i=0}^{N-1}\boldsymbol{c}_{i}f_{i}^{eq},
\end{align}
\begin{align} 
\rho c_{s}^{2}\boldsymbol{I}+\rho\boldsymbol{u}\boldsymbol{u}=\sum_{i=0}^{N-1}\boldsymbol{c}_{i}\boldsymbol{c}_{i}f_{i}^{eq}.  
\end{align}
\label{eq:MomentsEqLBM}
\end{subequations}

The simplest collision operator in the LBM framework is the Bhatnagar-Gross-Krook (BGK) model \citep{BGK_original_paper},

\begin{equation}
\Omega_{i}\left(\boldsymbol{f}\right)=-\frac{1}{\tau}\left(f_{i}-f_{i}^{eq}\right),\label{eq:BGKOperator}
\end{equation}

\noindent which was first proposed as a simplification of the continuous Boltzmann equation more than 30 years earlier than the emergence of the LBM. The monotonic relaxation time, $\tau$, is linked to the kinematic viscosity of the fluid, 

\begin{equation}
\nu=c_{s}^{2}\left(\tau-\frac{1}{2}\right).
\label{eq:tau_visc}
\end{equation}

Another form of collision operator that was also suggested before the origin of the LBM is the multiple-relaxation-time (MRT) scheme \citep{Cercignani1988,MRT2002}. In this case, the relaxation is performed in the space of hydrodynamic moments rather than the
distribution functions, with a transformation matrix, $\boldsymbol{M}$,
converting the set of populations into the set of moments, $\boldsymbol{m}=\boldsymbol{M}\boldsymbol{f}$. Each moment is relaxed with its own relaxation rate towards its equilibrium state, which is also a transformation of the population equilibrium state, $\boldsymbol{m}^{eq}=\boldsymbol{M}\boldsymbol{f}^{eq}$. The MRT-LBE then takes the form,

\begin{equation}
f_{i}(t+\Delta t,\boldsymbol{x}_{j}+\boldsymbol{c}_{i}\Delta t)-f_{i}(t,\boldsymbol{x}_{j})=-\sum_{j=0}^{N-1}\mathit{\left(\boldsymbol{M}^{-1}\boldsymbol{SM}\right)_{ij}}\left(f_{j}-f_{j}^{eq}\right),
\label{eq:MRT-LBE}
\end{equation}

\noindent in which the matrix $\boldsymbol{S}=diag\left\{ s_{0},s_{1,}\ldots,s_{N-1}\right\}$ contains all relaxation rates, some of which coincide with $\tau$ to recover the correct hydrodynamic viscosity. In fact, the BGK operator is a special case of the MRT operator with all rates equal to $\tau$. The superiority of the MRT-LBE over its BGK counterpart has been discussed in recent years with more advanced operators also proposed \citep{Geier2015_cumulant_LBM}.

One particular property of the MRT operator, namely the ability to control some spurious effects induced by bounce-back boundary conditions and non-hydrodynamic modes, favours its use in the simulation of viscoplastic fluids. This control is provided by the so-called \textit{magic numbers} in conjunction with the two-relaxation-time (TRT) scheme \citep{MagickNumbers2009}. These \textit{magic numbers} are specific combinations of the relaxation rates for the odd- and even-order modes, $\Lambda=\left(1/2-\tau_{odd}\right)\left(1/2-\tau_{even}\right)$, with $\tau_{even}=\tau$ connected to kinematic viscosity. The TRT model with $\Lambda=1/4$ is used throughout this work because it recovers the exact position of the wall in the case of the bounce-back boundary condition applied on a straight line.

In the context of the presented collision operators (i.e. BGK, MRT, TRT), the macroscopic properties of the medium emerge as a consequence of the mesoscale interaction of the distribution functions. However, in the LBM there exist constructions that focus on obtaining the desirable macroscopic behaviour for the macroscopic quantities and using this information to prescribe the underlying distribution functions \citep{Vikhansky2008_short_paper}. This type of operator is exploited in the IR model of yield-stress fluids investigated in this study. In order to comprehensively introduce this model, a more detailed discussion of how fluid stress and strain rate tensors arise in the LBM dynamics is necessary.

\subsection{Strain rate and deviatoric stress tensors in the LBM}

In this section distinction is made between the continuous ($f_{i}$) and discrete populations ($\bar{f}_{i}$) of the LBE. As already mentioned, the LBE is equivalent to the hierarchy of transport equations of the consecutive moments of the distribution functions. Using the definitions of the particle population moments in Equation \ref{eq:MomentsLBM} it can be shown that the zeroth moment of the LBE gives rise to the continuity equation, \citep{Dellar_2003_MRT_convergence},

\begin{equation}
\frac{\partial}{\partial t}\sum_{i=0}^{N-1}f_{i}+\nabla\cdot\sum_{i=0}^{N-1}\boldsymbol{c}_{i}f_{i}=0\;\Rightarrow\;\frac{\partial}{\partial t}\rho+\nabla\cdot\left(\rho\boldsymbol{u}\right)=0,
\label{eq:0_moment}
\end{equation}

\noindent noting that $\boldsymbol{c}_{i}\cdot\nabla f_{i}=\nabla\cdot\left(\boldsymbol{c}_{i}f_{i}\right)$. The transport equation for the first moment yields,

\begin{equation}
\frac{\partial}{\partial t}\sum_{i=0}^{N-1}\boldsymbol{c}_{i}f_{i}+\nabla\cdot\sum_{i=0}^{N-1}\boldsymbol{c}_{i}\boldsymbol{c}_{i}f_{i}=0\;\Rightarrow\;\frac{\partial}{\partial t}\left(\rho\boldsymbol{u}\right)+\nabla\cdot\sum_{i=0}^{N-1}\boldsymbol{c}_{i}\boldsymbol{c}_{i}f_{i}^{eq}=-\nabla\cdot\sum_{i=0}^{N-1}\boldsymbol{c}_{i}\boldsymbol{c}_{i}f_{i}^{neq},
\label{eq:1_moment_a}
\end{equation}

\noindent in which the equilibrium and non-equilibrium part, $f_{i}^{neq}=f_{i}-f_{i}^{eq}$, of the populations have been separated deliberately. The use of the properties
from Equation \ref{eq:MomentsEqLBM} reveals that Equation \ref{eq:1_moment_a} is in fact the momentum equation,

\begin{equation}
\frac{\partial}{\partial t}\rho u+\nabla\cdot\left(\rho c_{s}^{2}\boldsymbol{I}+\rho\boldsymbol{u}\boldsymbol{u}\right)=-\nabla\cdot\sum_{i=0}^{N-1}\boldsymbol{c}_{i}\boldsymbol{c}_{i}f_{i}^{neq},
\label{eq:1_moment_b}
\end{equation}

\noindent with the thermodynamic pressure,

\begin{equation}
p=\rho c_{s}^{2},
\label{eq:pressure}
\end{equation}

\noindent and, more importantly, the stress tensor,

\begin{equation}
\boldsymbol{\sigma}=-\sum_{i=0}^{N-1}\boldsymbol{c}_{i}\boldsymbol{c}_{i}\left(f_{i}-f_{i}^{eq}\right).\label{eq:Vikh_3}
\end{equation}

\noindent The transport equation for the second moment of the distribution function,
$\boldsymbol{\Pi}=\sum_{i=0}^{N-1}\boldsymbol{c}_{i}\boldsymbol{c}_{i}f_{i}$, then takes the form,

\begin{equation}
\frac{\partial}{\partial t}\boldsymbol{\Pi}+\nabla\cdot\left(\sum_{i=0}^{N-1}\boldsymbol{c}_{i}\boldsymbol{c}_{i}\boldsymbol{c}_{i}f_{i}\right)=\sum_{i=0}^{N-1}\boldsymbol{c}_{i}\boldsymbol{c}_{i}\Omega_{i}.
\label{eq:2nd_moment_base}
\end{equation}

The relations above are subjected to the formal small-scale parameter Chapman-Enskog expansion of the distribution function, $f_{i}=f_{i}^{\left(0\right)}+\epsilon f_{i}^{\left(1\right)}+\epsilon^{2}f_{i}^{\left(2\right)}+\ldots$, and time derivative, $\frac{\partial}{\partial t}=\frac{\partial}{\partial t_{0}}+\epsilon\frac{\partial}{\partial t_{1}}+\ldots$, with $f_{i}^{\left(0\right)}=f_{i}^{eq}$. This approach, together with the postulate of the MRT form for the collision operator, proves that the LBE reproduces the dynamics of the incompressible Navier-Stokes equations in the limit of $\epsilon\rightarrow0$ \citep{Dellar_2003_MRT_convergence}.

Following the transformation $f_{i}\rightarrow\bar{f_{i}}$ and use of the transport equations for the lower order moments, the fluid strain rate tensor can be linked with the second order tensor of the discrete non-equilibrium populations,

\begin{equation}
\boldsymbol{D}\approx-\frac{1}{2c_{s}^{2}}\sum_{i=0}^{N-1}\boldsymbol{c}_{i}\boldsymbol{c}_{i}\mathit{\mathit{\sum_{j=0}^{N-1}\mathit{\left(\boldsymbol{M}^{-1}\boldsymbol{SM}\right)_{ij}}\left(\bar{f_{j}}-f_{j}^{eq}\right)}}.\label{eq:LBM_MRT_strain_rate}
\end{equation}

The derivation of Equation \ref{eq:LBM_MRT_strain_rate} was performed for the general MRT collision operator. In the case of either the BGK or TRT operator, it takes a simpler form with only one relaxation rate present,

\begin{equation}
\boldsymbol{D}\approx-\frac{1}{2\tau c_{s}^{2}}\sum_{i=0}^{N-1}\boldsymbol{c}_{i}\boldsymbol{c}_{i}\mathit{\mathit{\left(\bar{f_{i}}-f_{i}^{eq}\right)}}.\label{eq:LBM_strain_rate}
\end{equation}

The work of \cite{Vikhansky2008_short_paper} took a different approach to linking the mesoscopic and macroscopic behaviour of the LBM. Although the Chapman-Enskog multiscale expansion is not used explicitly, the distribution function is still split into its equilibrium and non-equilibrium part, $f_{i}=f_{i}^{eq}+f_{i}^{neq}$, with an assumption that $f_{i}^{eq}\gg f_{i}^{neq}$. Thus the non-equilibrium component is neglected in the left hand side of Equation \ref{eq:2nd_moment_base},

\begin{equation}
\frac{\partial}{\partial t}\sum_{i=0}^{N-1}\boldsymbol{c}_{i}\boldsymbol{c}_{i}f_{i}^{eq}+\nabla\cdot\left(\sum_{i=0}^{N-1}\boldsymbol{c}_{i}\boldsymbol{c}_{i}\boldsymbol{c}_{i}f_{i}^{eq}\right)\approx\sum_{i=0}^{N-1}\boldsymbol{c}_{i}\boldsymbol{c}_{i}\Omega_{i},
\label{eq:Vikh_1}
\end{equation}

\noindent and the left hand side of Equation \ref{eq:Vikh_1} is shown to approximate the strain rate tensor,

\begin{equation}
\frac{\partial}{\partial t}\sum_{i=0}^{N-1}\boldsymbol{c}_{i}\boldsymbol{c}_{i}f_{i}^{eq}+\nabla\cdot\left(\sum_{i=0}^{N-1}\boldsymbol{c}_{i}\boldsymbol{c}_{i}\boldsymbol{c}_{i}f_{i}^{eq}\right)\approx2\rho c_{s}^{2}\boldsymbol{D}.\label{eq:Vikh_1andhalf}
\end{equation}

\noindent Thus, the second moment of the collision operator approximates the strain
rate tensor as well,

\begin{equation}
\mathbf{s}=\sum_{i=0}^{N-1}\boldsymbol{c}_{i}\boldsymbol{c}_{i}\Omega_{i}\approx2\rho c_{s}^{2}\boldsymbol{D}.
\label{eq:Vikh_2}
\end{equation}

Further details on the derivation of Equation \ref{eq:Vikh_2} can be found in \cite{Vikhansky_2011_Canada}. However, it is important to note here that the process exploits two key concepts, namely that the properties stated in Equation \ref{eq:MomentsEqLBM} introduce hydrodynamic variables in place of $f_{i}^{eq}$ while the transport Equations \ref{eq:0_moment} and \ref{eq:1_moment_b}
are used to replace the time derivatives with spatial ones.

\subsection{Implicit construction of the LBM}
\label{Vikh_LBM}

The central concept of the implicit construction of the LBM is to calculate the post-collision populations using the fluid stress and the second moment of the collision operator, $\boldsymbol{s}$,

\begin{equation}
\bar{f}_{i}^{pc}=f_{i}^{eq}+f_{i}^{\sigma}+\Phi_{i},\label{eq:Vikh_postCol}
\end{equation}

\noindent where $\Phi_{i}$ represents the implementation of the body force in this model and $f_{i}^{eq}$ is calculated one half-step back so that $\Phi_{i}/2$ must be subtracted from Equation \ref{eq:f_eq}. The distribution function $f_{i}^{\sigma}$ is specially crafted to carry the contribution of the stress and the second moment of the collision operator,

\begin{equation}
f_{i}^{\sigma}=\frac{3}{2}w_{i}\left(\boldsymbol{c}_{i}\boldsymbol{c}_{i}\right):\left(-\boldsymbol{\bar{\sigma}}+\frac{1}{2}\boldsymbol{s}\right),\label{eq:Vikh_f_stress}
\end{equation}

\noindent with only the deviatoric part of the stress included, $\bar{\boldsymbol{\sigma}}=\boldsymbol{\sigma}-Tr\left(\boldsymbol{\sigma}\right)\cdot\boldsymbol{I}$. Knowledge of the exact form of the collision operator, $\boldsymbol{\Omega}$, is unnecessary as only its second moment, $\boldsymbol{s}$, will be determined. More specifically, the following equation is exploited,

\begin{equation}
\frac{1}{2}\sum_{i=0}^{N-1}\boldsymbol{c}_{i}\boldsymbol{c}_{i}\Omega_{i}+\sum_{i=0}^{N-1}\boldsymbol{c}_{i}\boldsymbol{c}_{i}\left(\bar{f_{i}}-f_{i}^{eq}\right)=\sum_{i=0}^{N-1}\boldsymbol{c}_{i}\boldsymbol{c}_{i}\left(f_{i}-f_{i}^{eq}\right).\label{eq:Vikh_main}
\end{equation}

\noindent which is the second moment of Equation \ref{eq:pop_transform} after subtracting $f_{i}^{eq}$ from both sides. As it has been shown, the terms in Equation \ref{eq:Vikh_main} correspond to the fluid stress and strain. Thus this expression implicitly connects the second moment of the collision operator with the second moment of the transformed lattice populations and fluid stress (and, therefore, with strain rate, Equation \ref{eq:Vikh_2}),

\begin{equation}
\frac{1}{2}\boldsymbol{s}+\boldsymbol{T}=-\boldsymbol{\sigma},\label{eq:Vikh_main_short}
\end{equation}

\noindent where $\boldsymbol{T}=\sum_{i=0}^{N-1}\boldsymbol{c}_{i}\boldsymbol{c}_{i}\left(\bar{f_{i}}-f_{i}^{eq}\right)$. It is desirable to determine the stress and the second moment of the collision operator as the function of the known quantities, namely, $\boldsymbol{\sigma}=\boldsymbol{\sigma}\left(\boldsymbol{T}\right)$,
$\boldsymbol{s}=\boldsymbol{s}\left(\boldsymbol{T}\right)$. Subsequently, $f_{i}^{\sigma}$ can be calculated using Equation \ref{eq:Vikh_f_stress} and the particle populations updated using Equation \ref{eq:Vikh_postCol}.

In particular, when a rheological relation of the form $\boldsymbol{\sigma}=2\rho\nu\boldsymbol{D}$ is postulated, Equation \ref{eq:Vikh_main_short} yields the second moment of the collision operator and the stress tensor in the form,

\begin{equation}
\boldsymbol{s}=-\frac{1}{\frac{1}{2}+3\nu}\boldsymbol{T},\label{eq:Vikh_4}
\end{equation}

\begin{equation}
\boldsymbol{\sigma}=-\frac{3\nu}{\frac{1}{2}+3\nu}\boldsymbol{T}=2\rho\nu\cdot\left(-\frac{1}{2\rho c_{s}^{2}\tau}\right)\boldsymbol{T},\label{eq:Vikh_5}
\end{equation}

\noindent which coincide with the second moment of the BGK collision operator and the rheological relation of a Newtonian fluid with the strain rate in the form of Equation
\ref{eq:LBM_strain_rate}. In interpreting Equations \ref{eq:Vikh_4} and \ref{eq:Vikh_5} note that $\tau=1/2+3\nu$, as shown in Equation \ref{eq:tau_visc}. The procedure for solving Equation \ref{eq:Vikh_main_short} in case of the viscoplastic flow is outlined in Section \ref{sub:ImplicitRegularisedModel}.

\subsection{Implementation of body forces in the LBM}
\label{sub:BodyForce}

There exists several methods to introduce body force terms (e.g.
gravity, $g$) in LBM simulations. An extensive discussion of these methods can be found in \citep{Krafczyk2011_forces_review}. In this work, the formulation of \citep{Kupershtokh2009} known as the exact difference method (EDM) is used in conjunction with the TRT collision model. In the EDM the body force is introduced by adding
appropriate differences of the equilibrium distribution functions.
The LBE then takes the form,

\begin{equation}
f_{i}(t+\Delta t,\boldsymbol{x}_{j}+\boldsymbol{c}_{i}\Delta t)-f_{i}(t,\boldsymbol{x}_{j})=\Omega_{i}\left(f\right)+f_{i}^{eq}\left(\boldsymbol{u}+\boldsymbol{\Delta u}\right)-f_{i}^{eq}\left(\boldsymbol{u}\right).
\label{eq:14}
\end{equation}

Consecutive equilibrium distributions need to be calculated either
at the current velocity $\boldsymbol{u}$ or the one increased by $\boldsymbol{\Delta u}=\mathit{\boldsymbol{F}}\Delta t/\rho$. Additionally, the actual flow velocity is obtained after the addition of a correction term, $\mathbf{\mathit{\boldsymbol{u_{real}=u}}}+\boldsymbol{\Delta u}/2$.

The implicit construction of the LBM uses another method for body force implementation \citep{Vikhansky_2011_Canada} with the source term of the form,

\begin{equation}
\Phi_{i}=\frac{4w_{i}\left(\boldsymbol{c}_{i}\cdot\boldsymbol{F}\right)}{\rho},
\label{eq:Vikh_force}
\end{equation}

with the weights, $w_{i}$, coming from the equilibrium distribution.

\section{The LBM for generalised Newtonian fluids}
\label{sec:GenNewtonianLBM}

As shown in Equation \ref{eq:tau_visc}, the relaxation time in the LBM is directly linked to the fluid kinematic viscosity. Additionally, the rate of strain tensor can be obtained locally at each grid point using Equation \ref{eq:LBM_strain_rate}. It is therefore straightforward to implement non-Newtonian rheological behaviour by calculating the strain rate and then, in conjunction with a constitutive relation, adjusting the relaxation time. This
approach has been used extensively in simulations of various generalised-Newtonian fluids \citep{Leonardi2011_power_law} as well as in turbulence modelling using large eddy simulation \citep{TUB_LES_sphere_2011,MRT_LESvs_Cascaded_jet_2013,Suga2015_MRT_LES}.

The adjustment of the relaxation parameter can be employed in two ways. The first approach is explicit, meaning that the current viscosity is adjusted according to the rate of strain tensor computed using the relaxation time from the previous time step. In the second approach, a number of Newton iterations are performed in the collision process to determine the actual viscosity coincident with the rate of strain tensor. This implicit formulation is necessary because the expression for the rate of strain tensor (Equation \ref{eq:LBM_strain_rate}) is dependent on the relaxation time. In this work the explicit approach is used in the regularised model due to the fact that it was found to converge quickly to the correct steady-state solution.

Another approach to implementing strain-rate dependent viscosities in the LBM was reported by \cite{China_2008_rheology_in_FEQ} and \cite{Guo2015_nonNewtInFeq}. Here the non-Newtonian behaviour is provided by an additional term in the equilibrium distribution
functions, meaning that the particle distribution functions relax towards the desired rheology. Implementation of this term, however, requires non-local determination of the velocity gradients using finite difference stencils. This renders the approach less practical than those outlined in the following sections.

\subsection{Regularisation and relaxation time adjustment}
\label{sub:Regularised-model}

Explicit adjustment of the relaxation time can be used directly in the simulation of regularised Bingham plastics. The apparent viscosity is adjusted based on the local strain rate,

\begin{equation}
\nu_{app}\left(\dot{\gamma}\right)=\nu_{p}+\frac{\sigma_{y}}{\rho\dot{\gamma}}\left(1-\exp\left(-m\dot{\gamma}\right)\right),\label{eq:Pap_reg_1}
\end{equation}

\noindent where the rate of strain tensor is found from Equation \ref{eq:LBM_strain_rate}. Although the approach is conceptually straightforward, careful inspection reveals a numerical difficulty. In the case of exactly zero strain rate,
$\dot{\gamma}=0$, the numerical result becomes ill-posed
although analysis of the zero-strain behaviour shows that the limit is actually finite, $\lim\nu_{app}\left(\dot{\gamma}\right)_{\dot{\gamma}\to0}=\nu_{0}=\nu_{p}+m\sigma_{y}/\rho$. In particular, that means that numerical difficulties arise at the quiescent state. This issue is easily overcome by simple linear interpolation in the vicinity of the zero strain-rate,

\begin{equation}
\nu_{app}\left(\dot{\gamma}\right)=\frac{\dot{\gamma}^{C}-\dot{\gamma}}{\dot{\gamma}}\nu_{0}+\frac{\dot{\gamma}}{\dot{\gamma}^{C}}\nu_{app}\left(\dot{\gamma}^{C}\right),\label{eq:PR_interpolation}
\end{equation}
 
\noindent for $\dot{\gamma}<\dot{\gamma}^{C}$ where some small cut-off level of strain rate, $\dot{\gamma}^{C}$, is assumed. In this study it was sufficient to use $\dot{\gamma}^{C}=10^{-8}$ for calculations in double precision.

\subsection{Implicit regularisation for viscoplastic fluids} \label{sub:ImplicitRegularisedModel}

For the Bingham fluid rheology, Equation \ref{eq:Vikh_main_short} outlined in Section \ref{Vikh_LBM} can be solved analytically. In order to do that, one needs to consider unyielded and yielded regimes separately. 

In the first case, we simply postulate $\boldsymbol{s}=0$ that warranties essentially zero strain rate, see Equation \ref{eq:Vikh_2}. In this way, Equation \ref{eq:Vikh_main_short} immediately relates the stress tensor $\boldsymbol{\sigma}$ with the second moment of the non-equilibrium populations, $\boldsymbol{T}$. 

In the latter case, we use the rheological relation, Equation \ref{eq:rheoViscoplasticBase}. This time, however, the stress tensor is expressed as the function of the second moment, $\boldsymbol{s}$, instead of $\boldsymbol{D}$ (Equation \ref{eq:Vikh_2} used again),

\begin{equation}
\boldsymbol{\sigma}=\left(3\nu_{p}+\frac{\sqrt{2}\sigma_{y}}{s}\right)\boldsymbol{s},\label{eq:sigma_vs_s}
\end{equation}

with the tensor contraction $s=\sqrt{\boldsymbol{s}:\boldsymbol{s}}$. After combining Equation \ref{eq:sigma_vs_s} with Equation \ref{eq:Vikh_main_short} and contracting the result one gets the relation between the tensor contractions, $s\left(T\right)$ and $\sigma\left(T\right)$, $T=\sqrt{\boldsymbol{T}:\boldsymbol{T}},\:\sigma=\sqrt{\boldsymbol{\sigma}:\boldsymbol{\sigma}}$. To summarise, in both regimes the tensor contractions are given by,

\begin{equation}
\begin{cases}
\begin{array}{c}
\begin{array}{cc}
s=\frac{T-\sqrt{2}\sigma_{y}}{\frac{1}{2}+3\nu_{p}}, & \sigma=\frac{6\nu_{p}T+\sqrt{2}\sigma_{y}}{1+6\nu_{p}}\end{array}\\
\begin{array}{cc}
s=0, & \sigma=T,\end{array}
\end{array} & \begin{array}{c}
T\geq\sigma_{y}\\
T<\sigma_{y}
\end{array}.
\end{cases}
\label{eq:Vikh_sigmaS_formula}
\end{equation}

\noindent Thus, the tensors $\boldsymbol{\sigma}$ and $\boldsymbol{s}$ are as follows,

\begin{equation}
\begin{cases}
\begin{array}{c}
\begin{array}{cc}
\boldsymbol{s}=-\frac{T-\sqrt{2}\sigma_{y}}{\frac{1}{2}+3\nu_{p}}\cdot\frac{\boldsymbol{T}}{T}, & \boldsymbol{\sigma}=-\frac{6\nu_{p}T+\sqrt{2}\sigma_{y}}{1+6\nu_{p}}\cdot\frac{\boldsymbol{T}}{T}\end{array}\\
\begin{array}{cc}
\boldsymbol{s}=0, & \boldsymbol{\sigma}=-\boldsymbol{T},\end{array}
\end{array} & \begin{array}{c}
T\geq\sigma_{y}\\
T<\sigma_{y}
\end{array}.
\end{cases}
\label{eq:Vikh_sigmaS_tensor}
\end{equation}

\noindent Once $\boldsymbol{s}$ and $\boldsymbol{\sigma}$ are determined, the post-collision populations are found using Equation \ref{eq:Vikh_postCol}.

At this stage, some comments on the implicitly regularised viscoplastic model are in order. First, one can immediately notice that the process of \textit{implicit} regularisation takes place in the case of $T<\sigma_{y}$ and relies on the fact that the postulate that $\boldsymbol{s}=0$ is satisfied only approximately and the stress is exactly determined by the second moment $\boldsymbol{T}$. Second, it should be noted that Equation \ref{eq:Vikh_main_short} can be solved for a range of non-Newtonian constitutive relations. Bingham fluids admit for an analytical solution, but in principle many other fluid models could be solved numerically.

\subsection{Spurious terms in the rate of strain tensor} \label{sub:Burnett_theory}

As mentioned in Section \ref{sec:NavierStokesLBM}, the dynamics of the LBM is influenced by some terms from the underlying kinetic theory that are incompatible with purely hydrodynamic behaviour \citep{Dellar2014_Abstract2ndStress}. A careful inspection of the equation of transport of the deviatoric stress reveals
the so-called Burnett stress component, which is a $\mathcal{O}\left(Ma^{2}\right)$ diagonal term in the stress tensor. In the case of steady unidirectional channel flow in the x-direction, $\boldsymbol{u}=\left(u,0\right)$, this term is shown to be,

\begin{equation}
\sigma_{xx}^{Burnett}=-2\rho\nu\tau\left(\frac{\partial u}{\partial y}\right)^{2}.
\label{eq:BurnettStressPouisseile}
\end{equation}

The existence of this stress manifests itself in two ways. First,
when the deformation rate is computed using Equation \ref{eq:LBM_strain_rate}, the tensor $\boldsymbol{D}$ contains non-zero diagonal terms, $d_{xx}$, even in the case of zero longitudinal derivative, $\partial u/\partial x=0$,
and no density fluctuations. An example of this scenario is the case of periodic gravity-driven flow in a straight channel. Since the apparent viscosity in the Papanastasiou-regularised model exhibits very large rate of change in the close-to-zero strain
rate, the non-zero contribution from Equation \ref{eq:BurnettStressPouisseile} might have a significant effect on the value of viscosity. Second, it will influence the calculation of stresses in the implicitly regularised model by generating some spurious currents in the duct flow. The magnitude of this contribution is demonstrated in Section \ref{sub:Burnett_stress_results}.

\section{Results and discussion}
\label{sec:Results}

The convergence and performance of the PR and IR LBM models for for Bingham fluids is presented in this section. The comparative behaviour of the two formulations is highlighted using flows in circular and square ducts as well as creeping and inertial flows around a periodic array of spheres. However, prior to presenting these data, it is necessary to discuss the influence of the LBM lattice on the model results. 

\subsection{Influence of the lattice choice on LBM models}
\label{sec:LatticeChoice}

Lattice Boltzmann models can be formulated from a number of different velocity sets. While two-dimensional simulations almost always rely on nine velocities (i.e. the so-called D2Q9 model), the three-dimensional cases use various stencils. The models
D3Q15, D3Q19 and D3Q27 are most common, while some authors report using D3Q13 \citep{dHumieres_D3Q13_2001} or D3Q18 sets \citep{Vikhansky_2011_Canada}.

Only recently has it been reported that probably the most common model, D3Q19, produces some numerical artefacts in the form of spurious velocity currents in the case of axisymmetric channel flow configurations \citep{Texas_LBM_LES_spuriuos_2013,Suga2015_MRT_LES}.
These spurious currents are reinforced in turbulent flow simulations at sufficiently low values of viscosity. The detailed error analysis in the case of axisymmetric flows \citep{Suga2015_stencil_errors} showed that D3Q15 and D3Q19 models do not exhibit sufficient Galilean invariance.

\begin{figure}[!t]
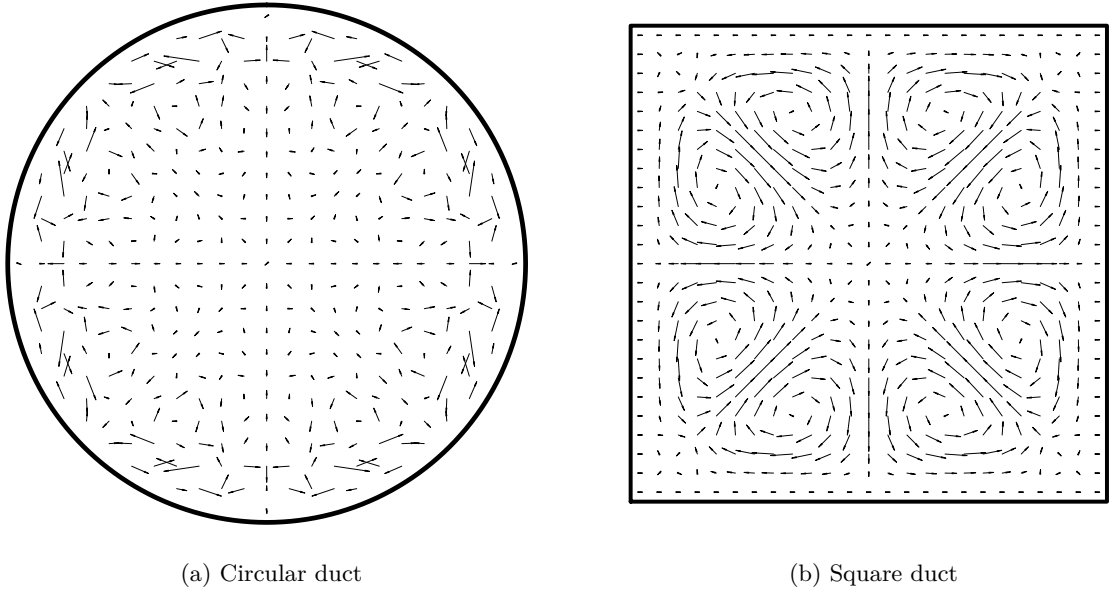

\subfloat[Circular duct]{
\input{spurious_currents_circle_D3Q19.tex}
}
\subfloat[Square duct]{
\input{spurious_currents_square_D3Q19.tex}
}

\caption{Spurious transverse currents in simple channel flows for the D3Q19 lattice stencil with constant viscosity.}
\label{fig:SpuriousCd3q19}
\end{figure}

In this work it is shown that it is not only the axisymmetric case of channel flow where spurious currents appear. The cases of both a circular and square duct under steady, unidirectional, laminar flow are shown to exhibit some regular patterns of those currents, as reproduced in Figure \ref{fig:SpuriousCd3q19}. In these models the lattice resolution was $25lu$. The maximal velocity for the circular duct flow was $U_{max}=0.01167$ while in the case of square duct it was $U_{max}=0.009734$, resulting in Mach numbers of $Ma=0.02$ and $Ma=0.017$, respectively. In both cases the magnitude of the spurious currents was $\mathcal{O}\left(10^{-7}lu/ts\right)$. These spurious currents were observed for both BGK and MRT models and were independent of the lattice resolution, as demonstrated in models using $100lu$.

In contrast, the D3Q27 stencil is completely devoid of these spurious effects and it is recommended that three-dimensional simulations are carried out using solely this model. This is the case in this work.

\subsection{Influence of the Burnett stress on viscoplastic fluids}
\label{sub:Burnett_stress_results}

As highlighted in Section \ref{sub:Burnett_theory}, the Burnett stress contribution in the LBM affects the deviatoric component of the fluid stress. Its influence on the simulation results is different for both the PR and IR viscoplastic models and the results are discussed as follows.

Unidirectional (in the x-direction), square-duct flows were investigated for both PR and IR models with a body force driving the flow. The results of the simulations were compared against the simulations where the diagonal components of the rate of strain and the stress tensors were set to zero, $d_{xx}=\sigma_{xx}=0$, in the LBM collision term.

\subsubsection{Papanastasiou-regularised model}
\label{subsec:Pap-reg-nu-dep}

According to Equation \ref{eq:Pap_reg_1}, the apparent viscosity should approach the limiting value,

\begin{equation}
\nu_{app}^{lim}=\nu_{p}+\frac{m\sigma_{y}}{\rho}.
\label{eq:nu_app_REG_lim}
\end{equation}

\begin{figure}
\hfill{}\input{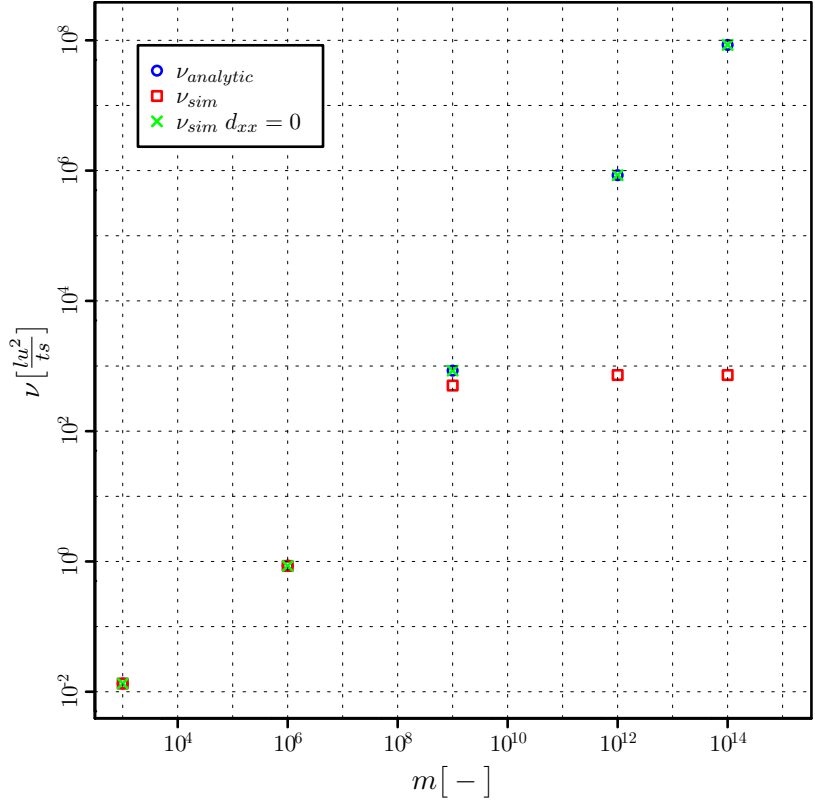}\hfill{}

\caption{Comparison of the simulation viscosity obtained using the PR model with that predicted analytically. The influence of the spurious Burnett stress terms becomes significant at the level of $m=10^{9}$, while the apparent viscosity saturates at the level of $\nu_{app}\approx727$ for higher regularisation parameters.}
\label{fig:nu_app_reg}
\end{figure}

\noindent It has been verified to what extent the value of $\nu_{app}$ is decreased, since it is expected that the strain rate is overestimated by the additional term from $\sigma_{xx}^{Burnett}$. Five values of regularisation parameters were investigated, namely $m=10^{3},\,10^{6},\,10^{9},\,10^{12},\,10^{14}$. The results are plotted in Figure \ref{fig:nu_app_reg} for the simulation parameters $\sigma_{y}=1.7\cdot10^{-6}$, $g=2.7\cdot10^{-6}$ and the height of the square duct, $H=63$. These results are compared against the simulation results for the standard model and the model with the spurious Burnett stress term $\sigma_{xx}=d_{xx}$ set to zero. It can be seen that the values of apparent viscosity diverge from $\nu_{app}^{lim}$ at approximately $m=10^{9}$ and saturate at a level which is orders of magnitude lower than that predicted by $\nu_{app}^{lim}$. For the artificially corrected strain-stress tensor, $d_{xx}=0$, the apparent viscosity coincides with the theoretically predicted value.

However, it was also found that the divergence of the value $\nu_{app}$ does not significantly influence the value of the maximal velocity in the channel. It is rather the value of the regularisation parameter itself that influences the accuracy of the solution. In the case of $m=1000$ the solution significantly departs from the expected flow behaviour simply because the contribution to viscosity is too low, namely $m\sigma_{y}/\rho\ll\nu_{p}$. As $m$ increases, good convergence is observed with the results for $m=10^{12}$ and $10^{14}$ actually coincident with each other down to numerical accuracy ($10^{-16}$). These results are summarised in Table \ref{tab:Rel_error_m}.

\begin{table}
\begin{tabular}{|c|c|}
\hline 
{\footnotesize{}Regularisation parameter $m$ }  & {\footnotesize{}Relative error $\varepsilon=\left|\frac{U_{max}-U_{max}\left(m=10^{14}\right)}{U_{max}\left(m=10^{14}\right)}\right|$}\tabularnewline
\hline 
\hline 
{\footnotesize{}$10^{3}$ }  & {\footnotesize{}$1.37$}\tabularnewline
\hline 
{\footnotesize{}$10^{6}$ }  & {\footnotesize{}$1.03\cdot10^{-2}$}\tabularnewline
\hline 
{\footnotesize{}$10^{9}$ }  & {\footnotesize{}$8.5\cdot10^{-6}$}\tabularnewline
\hline 
{\footnotesize{}$10^{12}$ }  & {\footnotesize{}$0$}\tabularnewline
\hline 
\end{tabular}

{\footnotesize{}\caption{\label{tab:Rel_error_m}Relative error in maximum velocity for the PR model where the reference datum is taken as $U_{max}\left(m=10^{14}\right)$.}
}{\footnotesize \par}
\end{table}

\subsubsection{Implicitly-regularised model}
\label{subsec:Imp-reg-nu-dep}

\begin{figure}[!t]
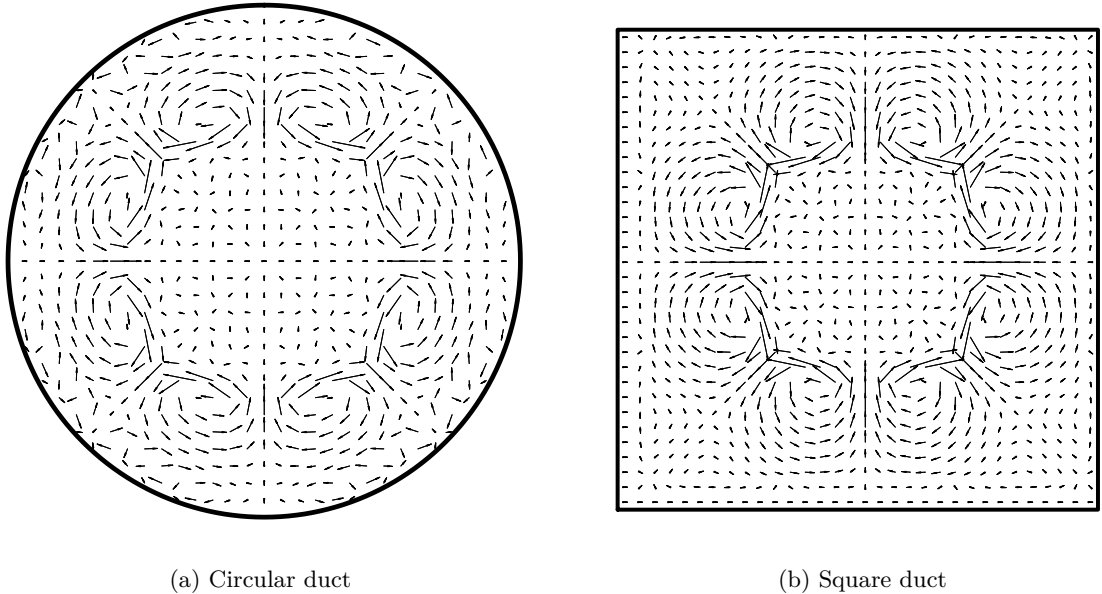

\subfloat[Circular duct]{
\input{spurious_currents_circle_VIKH.tex}
}
\subfloat[Square duct]{
\input{spurious_currents_square_VIKH.tex}
}

\caption{Transverse spurious currents in the viscoplastic IR model in the case of unidirectional duct flow, showing that their magnitude is greatest close to the interface of the yielded and unyielded fluid.}
\label{fig:Vikh_spurious_currents_Burnett}
\end{figure}

In the case of the implicitly regularised model, the additional contribution
of the Burnett term manifested itself in the presence of the spurious currents in the direction transverse to the main flow direction. In the case of unidirectional flow in the x-direction this phenomenon is easily explicable. During the deviatoric stress computation the non-zero term, $T_{xx}$, from the $\boldsymbol{T}$ tensor diagonal is redistributed to other flow directions, $T_{yy}$ and $T_{zz}$. This creates nonphysical loads that drive the flow. Some results are shown in Figure \ref{fig:Vikh_spurious_currents_Burnett}, where it can be seen that the magnitude of the currents is greatest close to the border of the unyielded surface. Their magnitude reaches approximately $3\cdot10^{-7}lu/ts$ for the circular case and $6\cdot10^{-7}lu/ts$ for the square case and is several orders of magnitude smaller than the magnitude of the their respective main flow velocities, $U_{max}=1.1\cdot10^{-2}$ and $U_{max}=1.92\cdot10^{-2}$.

The benchmark simulation with the term $T_{xx}$ set to zero was devoid of any spurious currents. The influence of those currents on the maximal velocity in the flow was negligible. In this study the relative difference of the maximal velocities was found to be $\mathcal{O}\left(10^{-7}\right)$. However, it must be noted that this comparison was possible only because the direction of the flow, and the associated stress components, were known \textit{apriori}. A general strategy for the removal of Burnett stresses is desirable and, as yet, unknown.

\subsection{Convergence in duct flows}
\label{sub:Convergence for duct flows}

The comparative convergence of the PR and IR viscoplastic models was investigated using two configurations. In these tests, the viscoplastic fluids was driven by a body force through ducts with both circular and square cross section. To create a control data set, the convergence behaviour of both approaches was also tested for the case of a purely Newtonian fluid. The relative error was defined with respect to the analytic solutions available, $\varepsilon_{rel}=\left|U_{max}^{sim}-U_{max}\right|/U_{max}$. The domain size employed, $D$, varied from 15 to 511 nodes with decreasing body force intensity, $g$, and fixed plastic viscosity, $\nu_{p}$. Since all cases need to be dynamically similar, the appropriate similarity numbers, namely the Hagen number, $Hg=gD^{3}/\mu^{2}$, and the Bingham number, $Bi=4\sigma_{y}/Dg$, were kept fixed. This implied the scaling $g\sim1/D^{3}$ and $\sigma_{y}\sim1/D^{2}$. The no-slip boundary condition on the walls was imposed using the standard bounce-back boundary condition.

\subsubsection{Analytical solutions for the duct flows}
\label{sub:AnalSols}

The maximal velocity of a Bingham fluid in circular duct flow can be expressed analytically,

\begin{equation}
\ensuremath{U_{max}=\frac{gR^{2}}{4\nu_{p}}\left(1-\frac{r_{y}}{R}\right)^{2},}
\label{eq:U_max_circle}
\end{equation}

\noindent where $R$ is the channel radius and $r_{y}=2\sigma_{y}/\rho g$ is the so-called yield radius, below which the fluid is not deformed. In addition, the expression for the maximal velocity in a square duct was derived using the variational approach \citep{Mosolov_Miasnikov_1966_limit_yield_surface_pipe_1} and can be written as,

\begin{equation}
\ensuremath{U_{max}=\frac{gc_{max}H^{2}}{16\mu_{p}}\left(1-\frac{Bi}{Bi_{cr}}\right)^{2},}
\label{eq:U_max_square}
\end{equation}

with the coefficient $c_{max}=1.178741652$, the Bingham number defined as $Bi=4\sigma_{y}/gH$, and the critical (i.e. maximal before the flow jams) Bingham number defined as $Bi_{cr}=4/\left(2+\sqrt{\pi}\right)\approx1.060318$. Various authors report slightly different values of $Bi_{cr}$, depending on the approach used. For example,  \citet{TaylorWilsonBicr112Reg} report $Bi_{cr}=1.12$ using a regularised model, \citet{Vikhansky2008_short_paper} finds the maximal value to be $Bi_{cr}\approx1.05$, and \citet{Roquet_Saramito_2001_aFEM_pipe} report $Bi_{cr}=1.07$ by extrapolation of FEM simulation data.

Initial investigations in this work showed that for all lattice resolutions the relative error for the flow in the square duct remained at 2.2\% for both the PR and IR model. This error was calculated with respect to Equation \ref{eq:U_max_square} and the originally proposed $Bi_{cr}=1.060318$. This result suggested that the semi-analytical relation is not sufficiently accurate and that another approach to estimate the rate of convergence for the square geometry was necessary. However, Equation \ref{eq:U_max_square} was found to be very accurate for the purely Newtonian case ($Bi=0$).

\subsubsection{Spectral solution for the duct flows}

To assist with the convergence study, a collocation-based spectral code \citep{CanutoQuarteroni2006Spectral} was written to solve the regularised version of the equations in the case of square duct flow. Papanastasiou-type regularisation yields very similar results to the implicitly-regularised model in the case of unidirectional duct flows but, from the point of view of finite element methods, it is more straightforward to implement due to the lack of a need to make a distinction between the yielded and unyielded regime.

The problem comes reduces to solving the nonlinear Poisson equation in a square domain with homogeneous (i.e. no-slip) boundary conditions,

\begin{equation}
\begin{cases}
\begin{array}{c}
\frac{\partial}{\partial x}\left(\nu_{app}\cdot\frac{\partial u}{\partial x}\right)+\frac{\partial}{\partial y}\left(\nu_{app}\cdot\frac{\partial u}{\partial y}\right)=-2q\\
u\left(x=-1,y\right)=u\left(x=1,y\right)=u\left(x,y=-1\right)=u\left(x,y=1\right)=0
\end{array} & \,\end{cases},
\label{eq:PoissonEq}
\end{equation}

\noindent where $q$ is the density-specific pressure gradient or bulk force intensity and apparent viscosity, $\nu_{app}=\nu_{app}\left(\partial u/\partial x,\partial u/\partial y,\nu_{p},\sigma_{y}/\rho,m\right)$,
is given by Equation \ref{eq:Pap_reg_1}.

The Gauss-Legendre-Lobatto (GLL) grid on the $\left[-1,1\right]$ section is used. The set of the grid nodes includes the endpoints and $N-2$ internal nodes being the roots of of the $\left(N-2\right)th$ order Jacobi polynomial, $P_{N-2}^{\left(\alpha,\beta\right)}$, with $\alpha=\beta=1$. The grid is simply the Cartesian product of this set on two dimensions. The base functions are defined as interpolating polynomials for the grid nodes (i.e. $u\left(x,y\right)=\sum_{i}^{N}\sum_{j}^{N}A_{i,j}L_{i}\left(x\right)\times L_{j}\left(y\right)$). The derivatives in Equation \ref{eq:PoissonEq} are calculated as the matrix-vector product of the function value in the grid nodes with the values of derivatives of the interpolating polynomials in those
nodes. The integration is carried out using GLL quadrature on the same grid. The discrete system of equations is solved using the DPOSV subroutine from LAPACK.

The external convergence loop is based on the direct iteration method in which  viscosity $\nu_{app}$ is calculated using the coefficients and velocity derivatives after the system has been solved and is treated as input for the next iteration step. The solution procedure is terminated based on the criterion of maximum relative change of the nodal values in two consecutive time steps.

Since viscosity gradients are large, the system becomes stiff from the numerical point of view and it is difficult to obtain good convergence for large values of $m$. The reference solution was calculated for the regularisation parameter $m=10^{8}$ in the square geometry and $N=129$ quadrature points along a single direction. The maximum velocity is $U_{max}=6.3121307$ with the relative error between two consecutive iterations at the level of $\varepsilon=10^{-4}$. Of course, the value of maximal velocity was re-scaled to non-dimensional numbers, resulting in $Re=12.6242614$ for $Hg=432$ and $Bi=0.4$.

\begin{figure}[!t]
\hfill{}\input{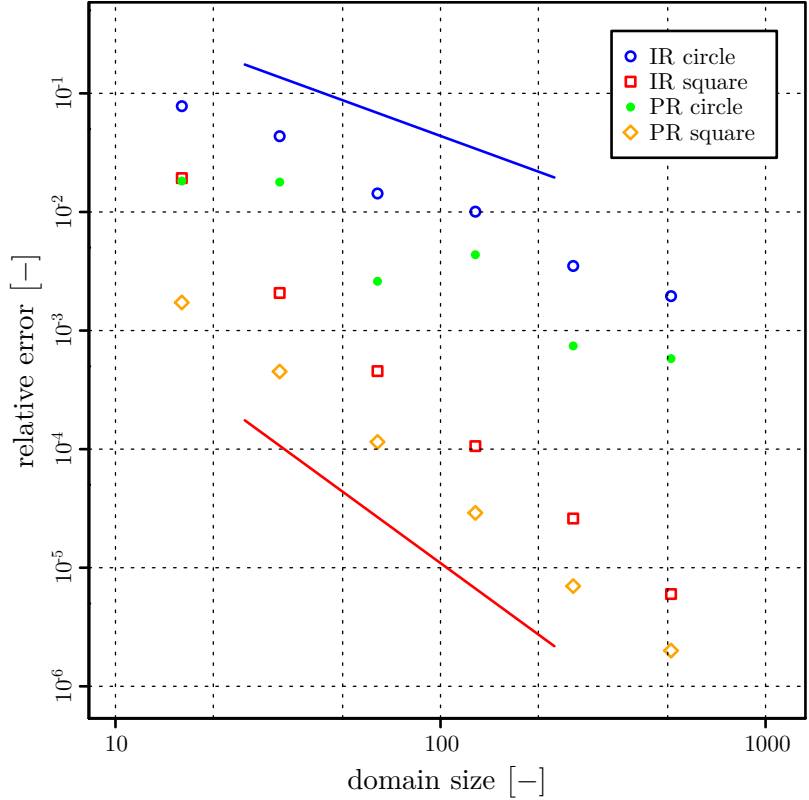}\hfill{}

\caption{Convergence rate for the Newtonian version of the PR and IR models in circular and square duct flows, with lines of slope -1 and -2 plotted for reference. The higher order of convergence for the square geometry suggests that it is only the boundary condition formulation that spoils the order in the case of circular duct.}
\label{fig:Conv-newtonian_3d}
\end{figure}

With the spectral solution for the square duct defined, the convergence study could continue. Figure \ref{fig:Conv-newtonian_3d} shows the convergence rate for the case of Newtonian fluids. In this graph the error is calculated relative to the analytical solutions in Equations \ref{eq:U_max_circle} and \ref{eq:U_max_square} and plotted as a function of the domain size. Lines with the slope of -1 and -2 are also plotted for reference. The simulations were performed at the same value of numerical viscosity, $\nu_{LB}=0.0125$ with decreasing body force and increasing resolution, $g=g_{0}\left(N_{0}/N\right)^{3}$, so as to keep the Hagen number,  $Hg=gN^{3}/\nu_{LB}^{2}$, fixed at a value of $432$. The reference values were $g_{0}=2\cdot10^{-5}$ and $D=H=15lu$ and the resolution was increased up to $511lu$ according to $D_{i}=2^{i+4}-1$, with the odd number of fluid nodes facilitating the probing of velocity on the centreline of the channel. In these results it can be seen that the flow in the square duct converges with second order accuracy while the case of circular duct is only first order accurate. The latter effect can be attributed to the bounce-back boundary condition that represents the geometry using a staircase approximation. Thus, the location of the wall is only first order accurate and this error dominates the accuracy of the simulation. In the case of the square geometry the location of the wall is known to reside close to half of the distance between two nodes.

\begin{figure}[!t]
\hfill{}\input{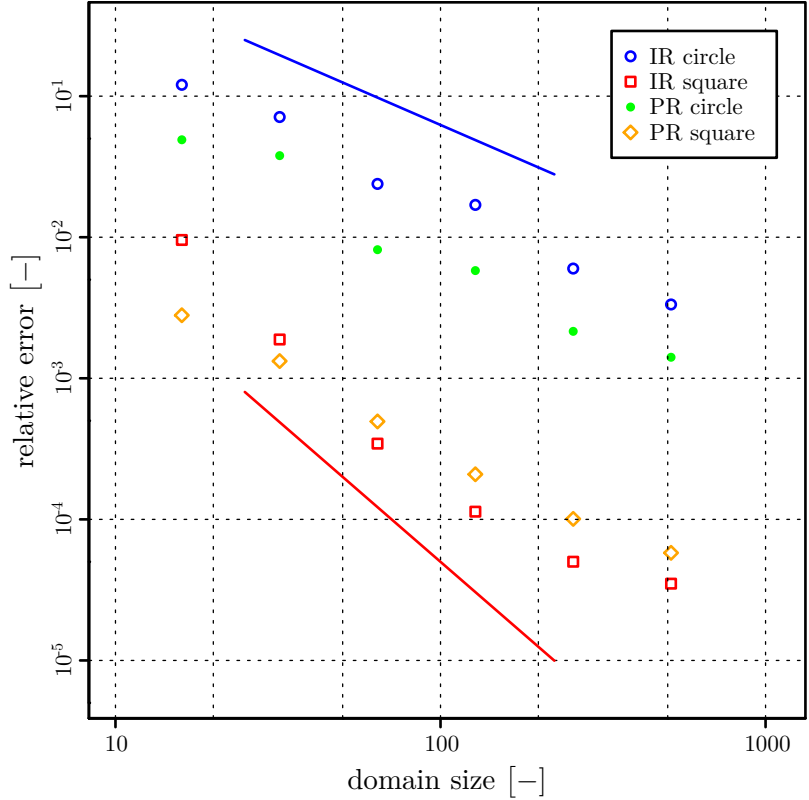}\hfill{}

\caption{Convergence rate for the viscoplastic version of the PR and IR models in circular and square duct flows, with lines of slope -1 and -2 plotted for reference. For the square duct, both the PR and IR models initially converge with second order accuracy until the slope of the convergence flattens out, possibly due to inaccuracies in the reference spectral solution.}
\label{fig:Conv-viscoPlastic_3d}
\end{figure}

The convergence results for the viscoplastic models are presented in Figure \ref{fig:Conv-viscoPlastic_3d}. The model configuration is the same as that described in Figure \ref{fig:Conv-newtonian_3d} except for the addition of the non-zero Bingham number, $Bn=4\sigma_{y}/\rho gD=0.4$. The yield stress was scaled as $\sigma_{y}=\sigma_{y0}\left(N_{0}/N\right)^{2}$ with $\sigma_{y0}=3\cdot10^{-5}$. Lines with the slope of -1 and -2 are again plotted for reference. Both formulations report first order accuracy for the circular geometry and almost second order accuracy for the flow in the square duct at coarse resolutions. For the square duct, the PR and IR models initially converge with second order accuracy until the slope of the convergence flattens out. For the square duct the error is calculated relative to the reference spectral solution which itself has some degree of inaccuracy, which may be the reason why the order of convergence decreases with increasing lattice resolution.

\subsection{Transient behaviour in duct flows}

Both the PR and IR viscoplastic models were also compared in transient flow cases. This was done using the case of impulsively-started quiescent flow in a square duct. Again, the flow is driven by a body force. It was expected that the fluid would accelerate with (theoretically) asymptotic approach to the maximal velocity. Since the flow is fully unyielded (i.e. rigid) at the initial condition, the unyielded zone shrinks while the central velocity increases. At the same time, however, it is expected that the unyielded zone exhibits a plateau in the velocity profile.

\begin{figure}[!t]
\subfloat[PR (left) vs. IR 15-100ts]{
\input{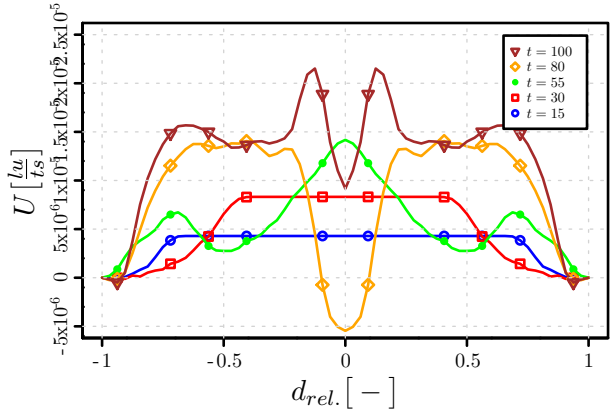}
\input{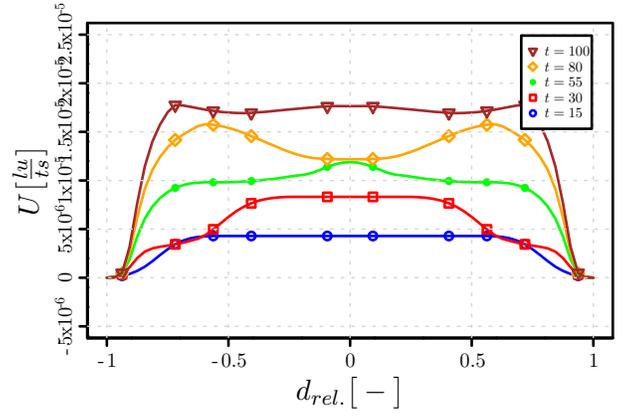}
}

\subfloat[PR (left) vs. IR 150-800ts]{
\input{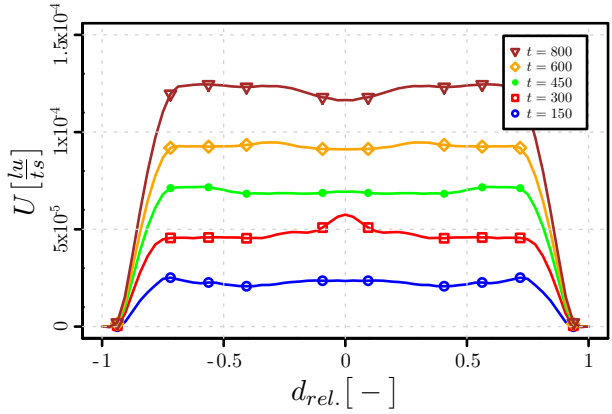}
\input{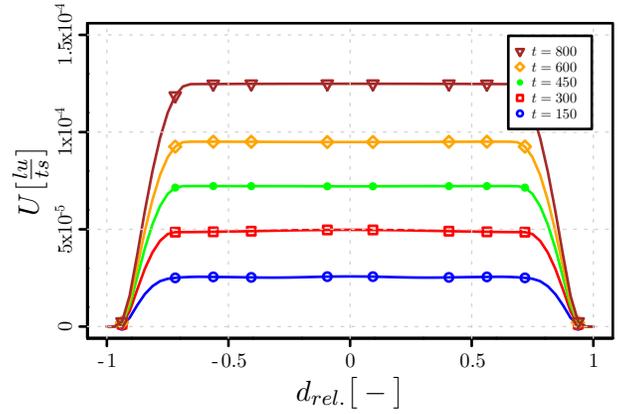}
}

\subfloat[PR (left) vs. IR 1000-2000ts]{
\input{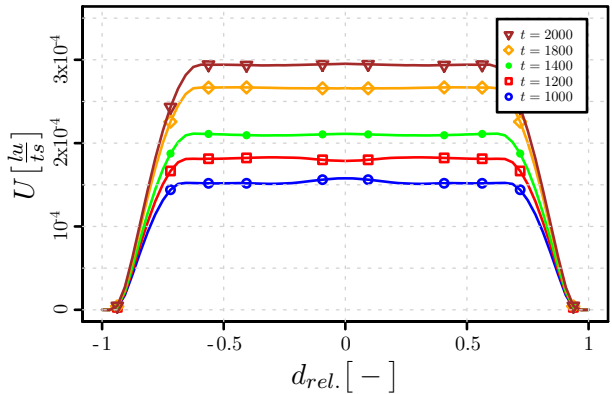}
\input{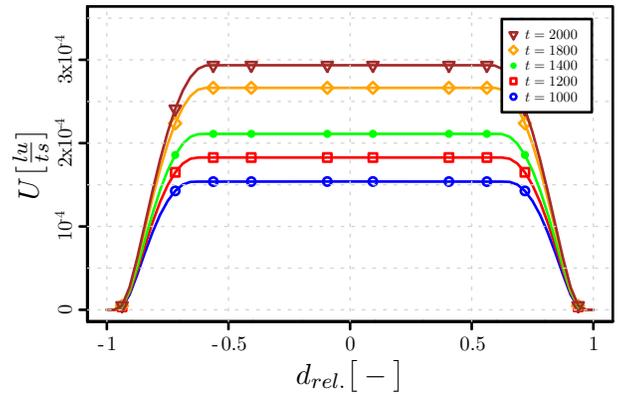}
}

\caption{Diagonal velocity profiles for the case of impulsively started flow in a square duct, showing that the PR approach exhibits strong nonphysical fluctuations shortly after the onset of motion and is clearly inferior to the IR model in this respect.}
\label{fig:Transient-behaviour-duct}
\end{figure}

The test geometry used in these investigations is a square duct that is 63 $lu$ wide. The velocity profile on the diagonal of the square during various phases of the flow acceleration is shown in Figure \ref{fig:Transient-behaviour-duct}. The flow configuration for these results is the same as in the case of the channel flow convergence study for the third lattice size. The  driving body force was $g=1.7\cdot10^{-6}lu/ts^{2}$, the fluid viscosity was $\nu=0.0125$, and the yield stress was $\sigma_{y}=2.7\cdot10^{-6}lu^{2}/ts^{2}$. The profiles were plotted at selected time steps in three time segments, namely $15-100ts$, $150-800ts$ and $1000-2000ts$. The number of time steps to $\mathcal{O}\left(10^{-10}\right)$ convergence of the maximal velocity, $U_{max}\approx2.5\cdot10^{-3}$, was found to be $\mathcal{O}\left(10^{6}\right)$. Consequently, the behaviour shown in these graphs is at short time scales. It can be seen that the PR approach exhibits strong nonphysical fluctuations shortly after the onset of motion and is clearly inferior to the IR model in this respect. These fluctuations result in reversed flow in the PR model which is undesirable. The IR model also exhibits spurious fluctuations however their magnitude is smaller and they are attenuated much more effectively.

\subsection{Flow past a periodic array of cylinders}

This section focuses on verification of both the PR and IR models in the case of flows past a periodic array of circular cylinders. This is effectively a two-dimensional flow configuration with a single cylinder of diameter $D=0.4\cdot L$, where $L$ is the size of the the cell. This results in a solid volume fraction of $\phi\approx0.1257$. The flow was investigated in both the creep and inertial regimes ($Re\sim1$ and $Re\sim150$ respectively). Additionally, in the case of the IR model, the flow is investigated for constant viscosity with varying solid volume fraction in the range $0.05-0.70$ where the drag is compared against other data available in the literature.

\subsubsection{Newtonian creeping flow}

A range of particle diameters was investigated and the friction coefficient,

\begin{equation}
k=\frac{F_{drag}}{4\pi\nu U},
\label{eq:PeriodicFriction}
\end{equation}

\noindent was compared against data reported in the literature \citep{AcrivosSangani1982,Williams_Holmes_SPH_porous_2011,Zhu_1999}. Note that in Equation \ref{eq:PeriodicFriction} the superficial velocity $U=u_{bulk}\cdot\left(1-\phi\right)$ in which $\phi$ is the solid volume fraction, $u_{bulk}=Q/L^{2}$ and $Q$ is the volumetric flow rate. The simulations were performed
on three different grid resolutions, namely $L=125, 250$, and $500$, with decreasing Mach number. The flows were driven by a body force, with the resulting drag force calculated from the integral of the body force, $F_{drag}=gL^{2}\left(1-\phi\right)$. The drag forces from each resolution were found to be in close agreement with each other, with relative differences from $0.2\%$ for $\phi=0.05$ to $2.5\%$ for $\phi=0.70$. The discrepancy at the level of $2.5\%$ may appear large but it must be remembered that the LBM faces some intrinsic difficulties in creeping flow simulations with the bounce-back boundary condition \citep{Ginzburg_2015_spheres}.

\begin{figure}
\hfill{}\input{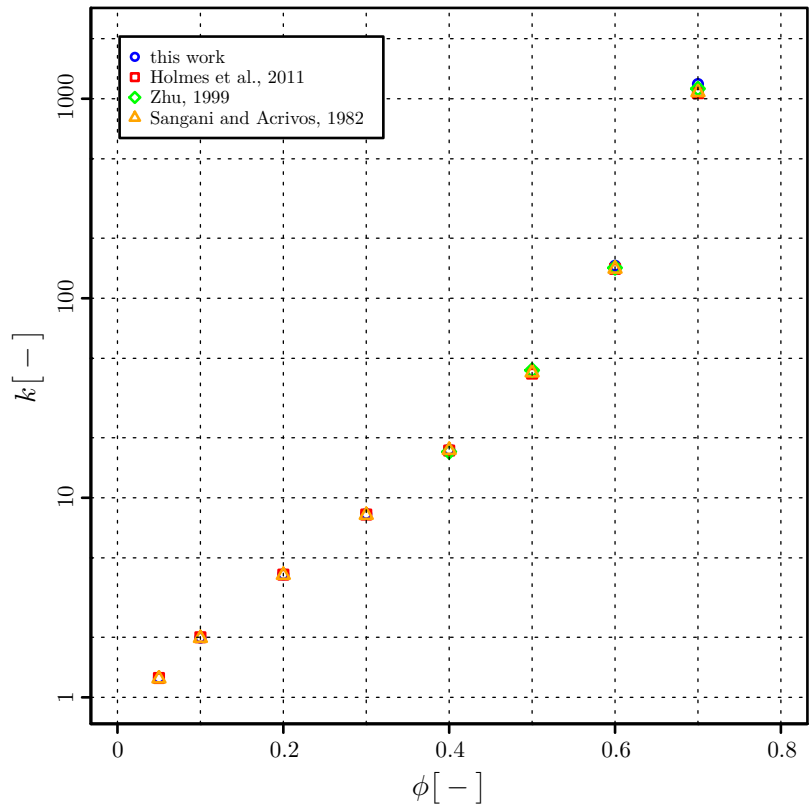}\hfill{}

\caption{Friction coefficient, $k$, calculated by the IR model for the case of Newtonian flow past a periodic array of cylinders. Comparison with data available in the literature at a range of solid volume fractions is also included.}
\label{fig:frict_coef_VIKH_NEWT}
\end{figure}

The Newtonian creeping flow data is presented in Figure \ref{fig:frict_coef_VIKH_NEWT}. In these results the IR model was employed with a Newtonian fluid, a grid resolution of $L=500lu$, viscosity $\nu_{LB}=0.5$, and driving body force intensity $g=9.8\cdot10^{-8}$. In all but one case the LBM results predict the largest value of the drag coefficient. Good agreement between all methods is observed in cases of small solid volume fractions ($\phi=0.05-0.4$) with discrepancies at the level of $1-2.5\%$. For larger solid volume fractions the LBM predicts the value of $k$ approximately $3-4\%$ larger than the other results and almost $10\%$ larger for the last case (i.e. $\phi=0.7$). Still, the fact that the discrepancy between the LBM and other approaches seems to show a regular pattern suggests that some intrinsic issues of the LBM (e.g. staircase geometry representation, relaxation rate-dependent position of the \textit{actual} boundary) can influence the results. Nevertheless, the accuracy of the method is considered here to be satisfactory.

\subsubsection{Viscoplastic creeping flow}

The PR and IR models were then compared in the viscoplastic creeping flow regime ($Re\sim1)$. The flow Reynolds number, $Re=U_{avg}D/\nu_{p}$, Bingham number, $Bn=\sigma_{y}D/\nu_{p}U_{avg}$, and friction factor, $f=2Dg/U_{avg}^{2}$, were compared to each other as well as to the results published by \citet{Vikhansky2008_short_paper}. The average velocity is defined as $U_{avg}=Q/L^{2}$. The flows were simulated on an $80\times80$ node grid with periodic boundary conditions and $D=0.4L$ resulting in a $32 lu$-wide cylinder in the centre. The driving body force was $g=1.25\cdot10^{-6}$, the plastic viscosity was $\nu_{p}=0.08$, and the yield stress, $\sigma_{y}$, was on the order of $10^{-6}$ with a single purely Newtonian case, $\sigma_{y}=0$. No mesh dependence study was carried out, since the implemented model follows that presented by \citet{Vikhansky2008_short_paper} and results reported in that work were verified with respect to mesh density and exactly the same resolution was used. The results of these simulations are summarised in Table \ref{tab:ParticleCreep}. Both models report good agreement, with discrepancy less than $1\%$ in every parameter except the relative difference of $2\%$ for the friction factor, $f$. A systematic discrepancy between the reference case is reported. The flow is slower in the reference case, which suggests that the size of the particle was slightly larger in that scenario.

\begin{figure}
\subfloat[$\sigma_{y}=0$]{
\input{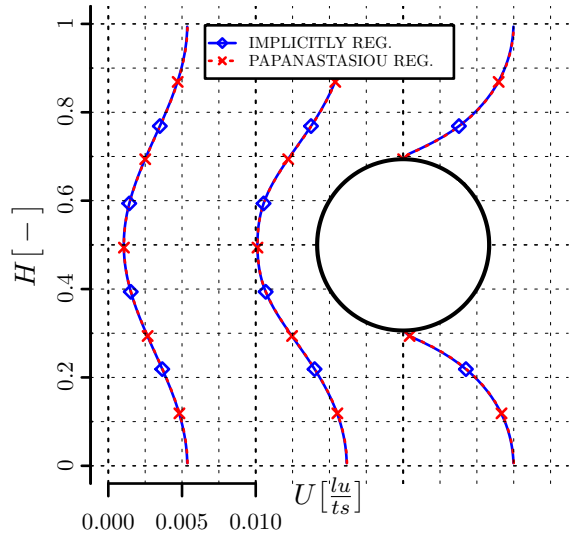}
}
\subfloat[$\sigma_{y}=10^{-6}$]{
\input{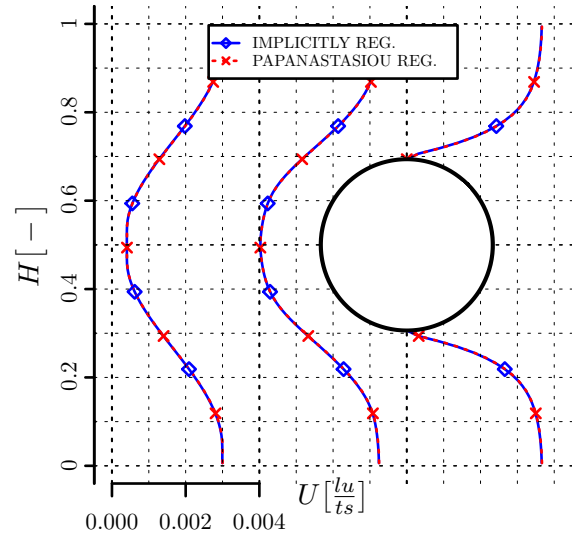}
}\\
\subfloat[$\sigma_{y}=1.5\cdot10^{-6}$]{
\input{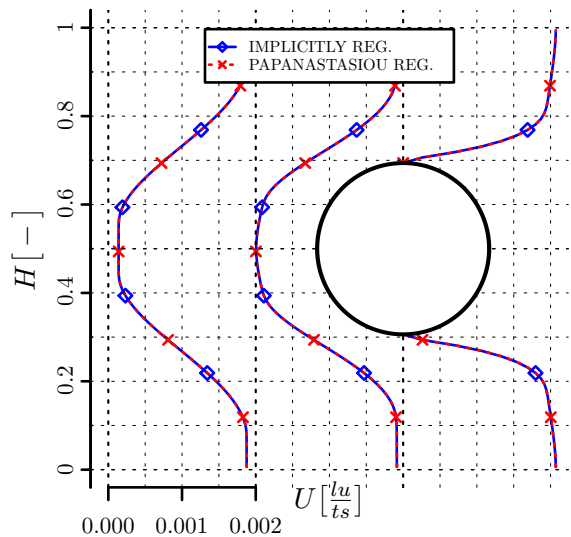}
}
\subfloat[$\sigma_{y}=2\cdot10^{-6}$]{
\input{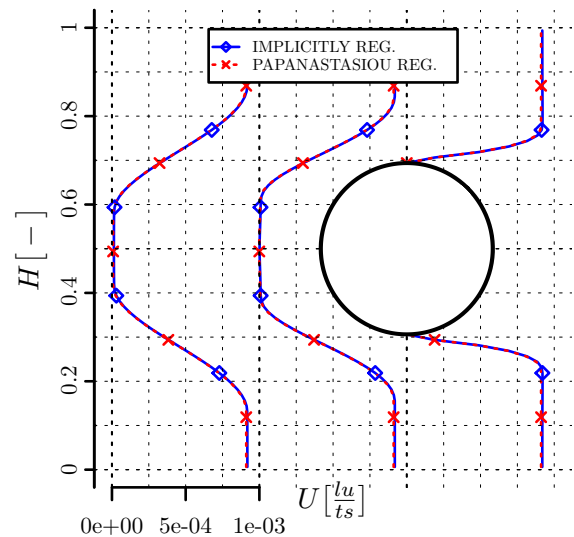}
}
\caption{Horizontal velocity profiles in the viscoplastic creeping flows past a periodic array of cylinders. Velocity profiles are plotted at the start, quarter and half of the cell. The PR and IR models exhibit very close agreement, with the only difference visible in the case of the the largest yield stress.}
\label{fig:CreepVelProfiles}
\end{figure}

Both approaches yield results which are close to each other in terms of the similarity numbers as well as the detailed velocity profiles, which are shown in Figure \ref{fig:CreepVelProfiles}. It must be highlighted that the numerical behaviour of the IR model was again found to be favourable. It reaches the steady state solution down to the numerical precision ($\mathcal{O}\left(10^{-16}\right)$) after $2\cdot10^{5}$ time steps. In contrast, the fluid velocity in the PR model only converges to the level of $10^{-12}lu/ts$ in the velocity field.

\begin{figure}
\subfloat[$\sigma_{y}=0$]{
\includegraphics[scale=0.30]{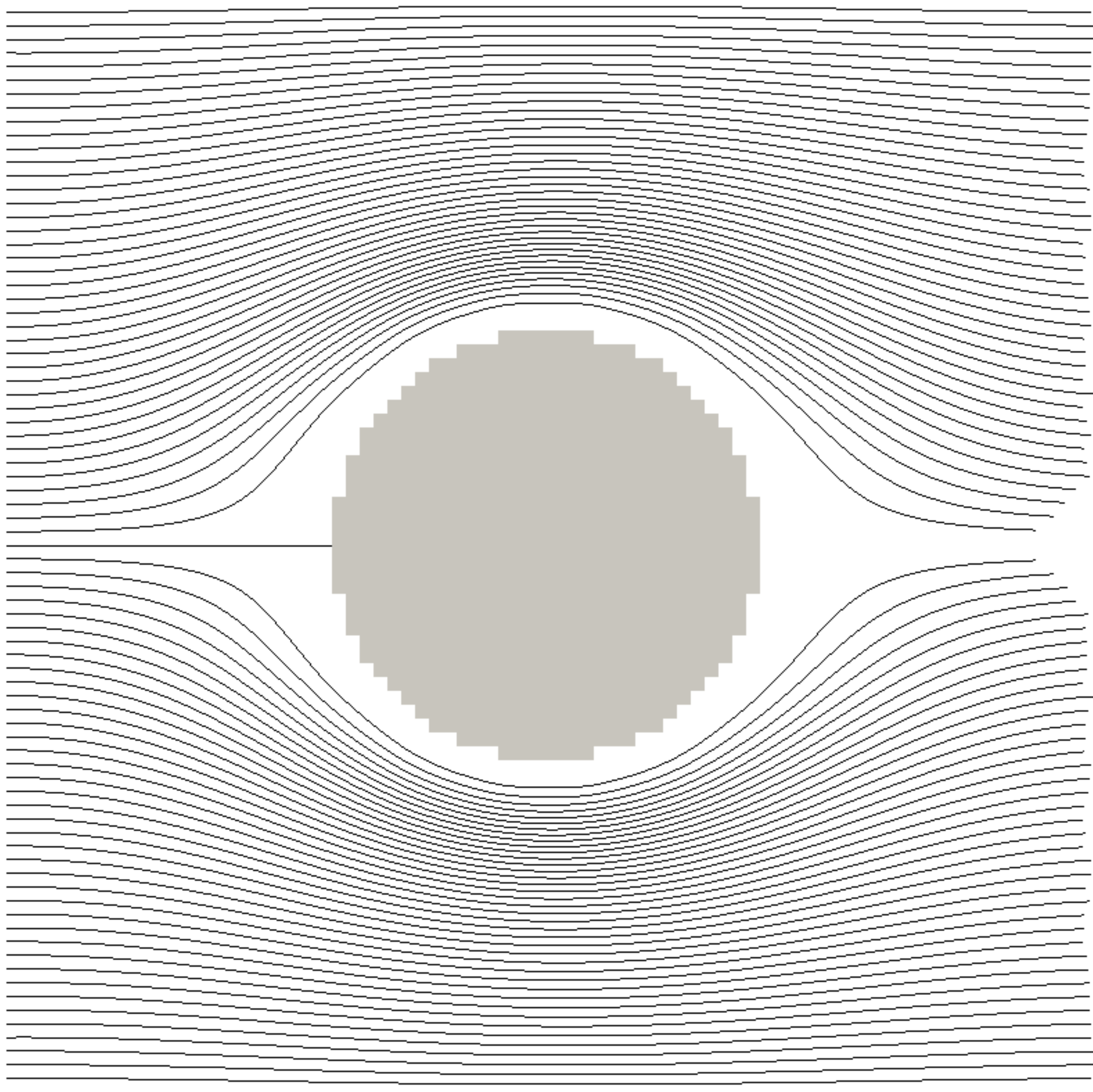}
}
\subfloat[$\sigma_{y}=10^{-6}$]{
\includegraphics[scale=0.30]{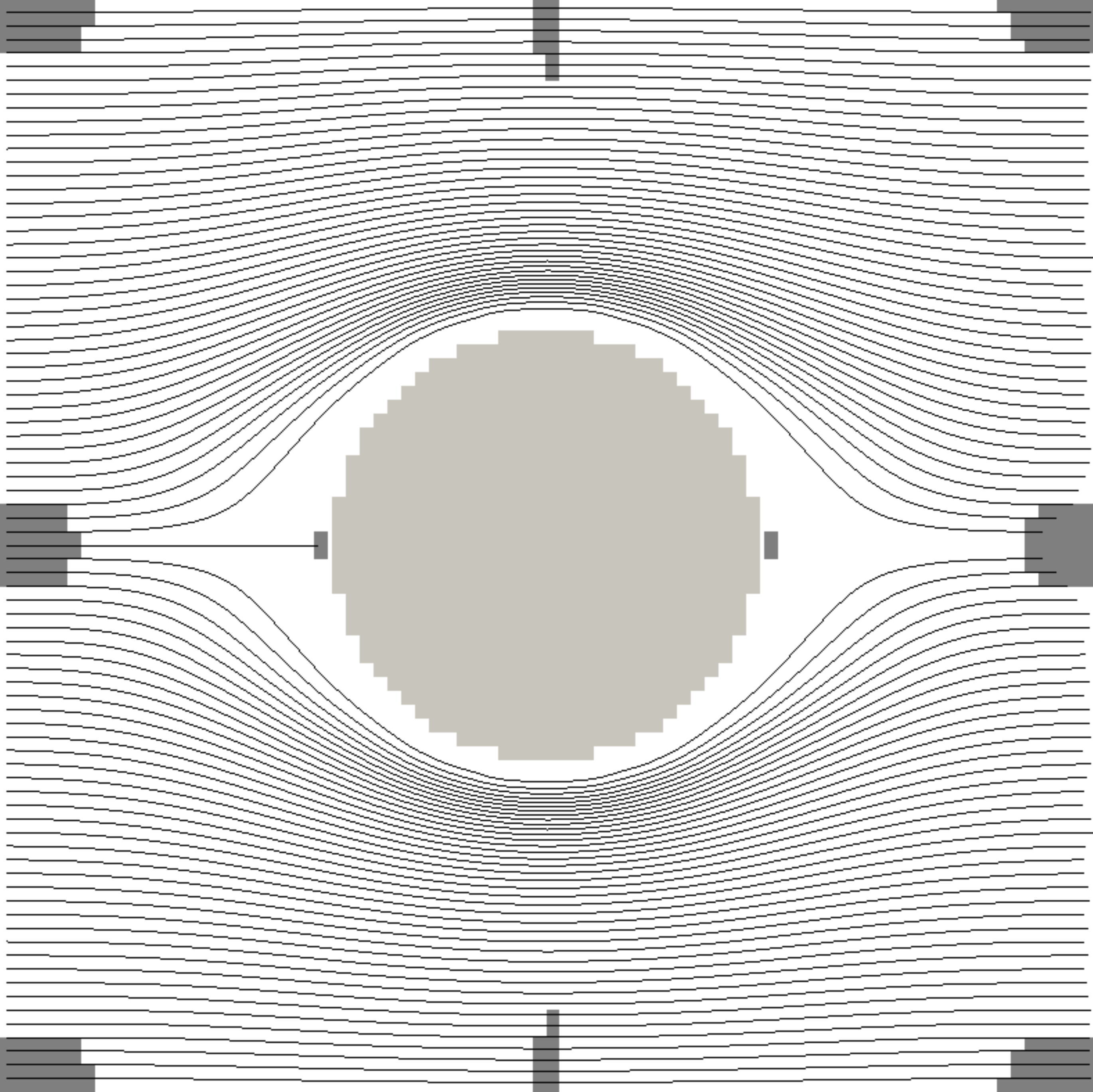}
}\\
\subfloat[$\sigma_{y}=1.5\cdot10^{-6}$]{
\includegraphics[scale=0.30]{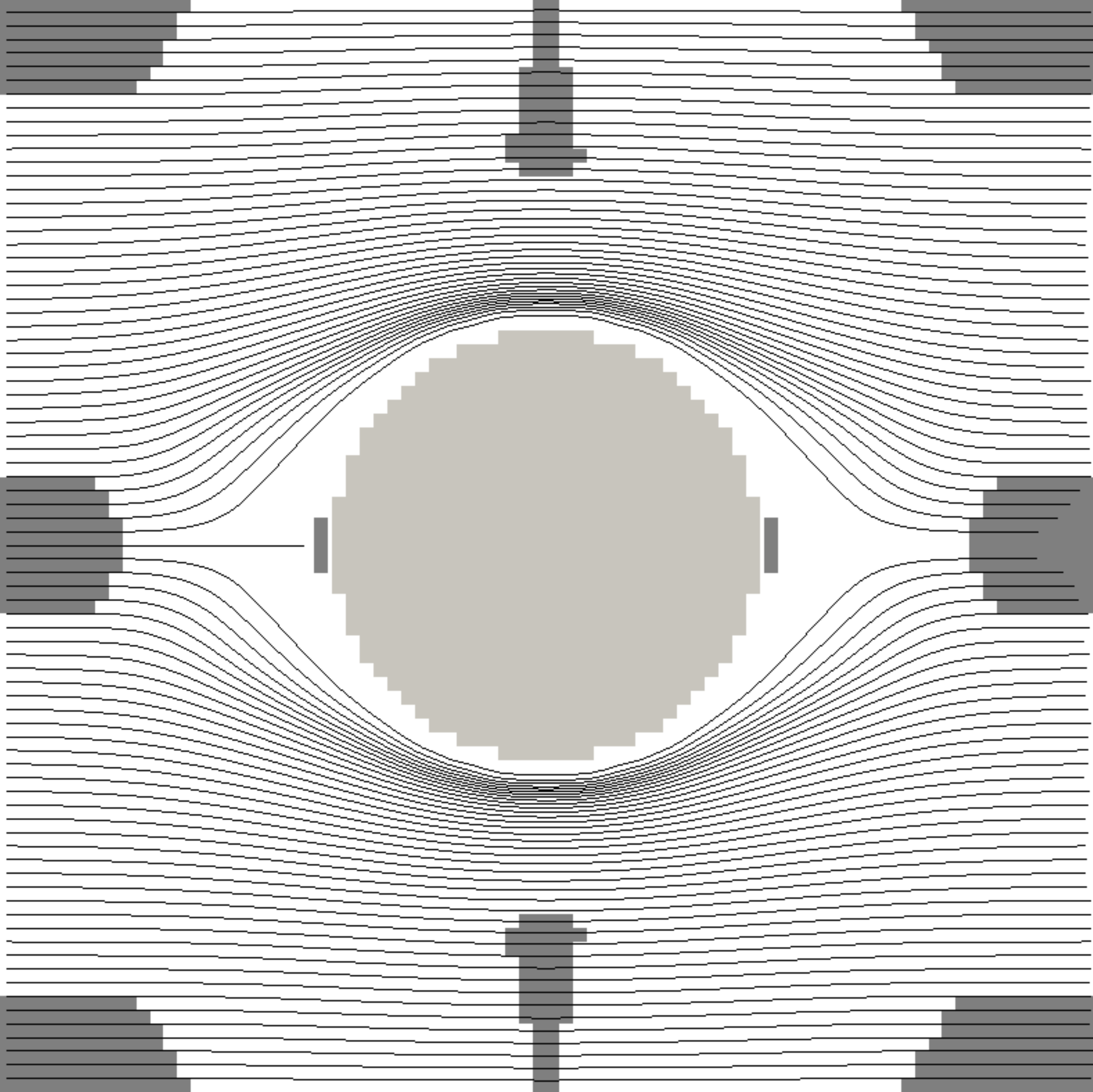}
}
\subfloat[$\sigma_{y}=2\cdot10^{-6}$]{
\includegraphics[scale=0.30]{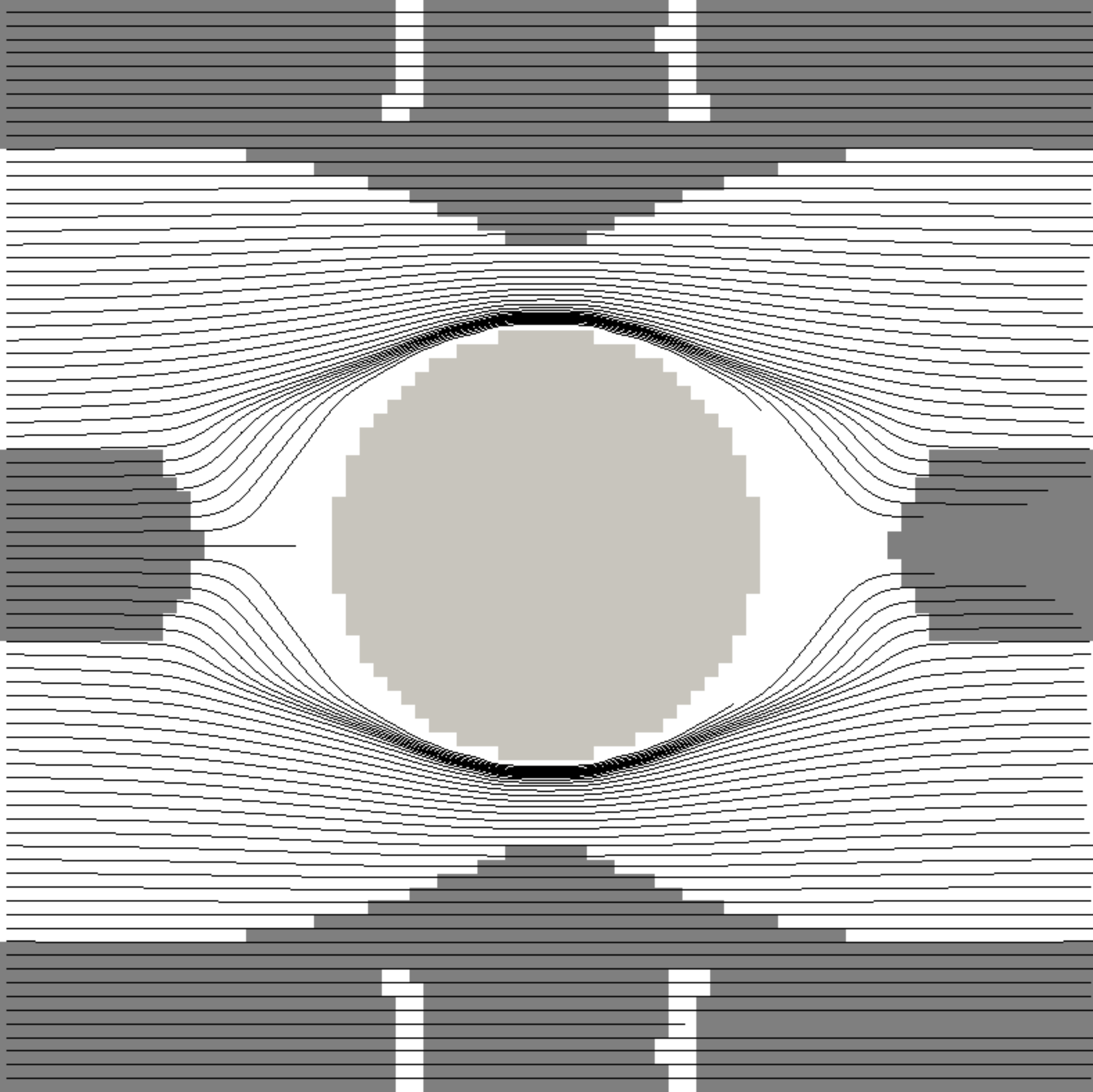}
}
\caption{The streamlines along with the yielded and unyielded zones for the IR model in the case of creeping flow past a periodic array of cylinders, showing the size of the unyielded zone increasing as the value of $\sigma_{y}$ increases.}
\label{fig:CreepStreamlines}
\end{figure}

In addition, the fluid streamlines for viscoplastic creeping flow past a periodic array of cylinders are included in Figure \ref{fig:CreepStreamlines}. It can be seen that the size of the unyielded zone increases as the value of $\sigma_{y}$ increases, and the resulting flow fields confirm the hypothesis that the particle size in the reference work \citep{Vikhansky2008_short_paper} is slightly larger. In that work the unyielded zones are smaller and the flow is slower. 

\begin{table}
\begin{tabular}{|c|c|c|c|c|c|c|c|c|c|}
\hline 
 & \multicolumn{3}{c|}{Implicitly reg. (IR)} & \multicolumn{3}{c|}{Papanastasiou reg. (PR)} & \multicolumn{3}{c|}{\citet{Vikhansky2008_short_paper} (IR)}\tabularnewline
\hline 
\hline 
$\sigma_{y}$  & $Re$  & $Bn$  & $f$  & $Re$  & $Bn$  & $f$  & $Re$  & $Bn$  & $f$\tabularnewline
\hline 
$0$  & $1.476$  & $0$  & $5.870$  & $1.483$  & $0$  & $5.818$  & $1.31$  & $0$  & $7.44$\tabularnewline
\hline 
$10^{-6}$  & $0.802$  & $1.994$  & $19.890$  & $0.805$  & $1.988$  & $19.764$  & $0.72$  & $2.22$  & $24.7$\tabularnewline
\hline 
$1.5\cdot10^{-6}$  & $0.486$  & $4.937$  & $54.163$  & $0.487$  & $4.933$  & $54.076$  & $0.444$  & $5.41$  & $64.9$\tabularnewline
\hline 
$2\cdot10^{-6}$  & $0.235$  & $13.606$  & $231.389$  & $0.233$  & $13.730$  & $235.643$  & $0.221$  & $14.5$  & $263$\tabularnewline
\hline 
\end{tabular}

\caption{Comparison of the similarity numbers in the case of creeping flow past a periodic array of cylinders in an $80\times80$ cell with a cylinder of $D=32lu$ and bounce-back boundary conditions.}
\label{tab:ParticleCreep}
\end{table}

\subsubsection{Viscoplastic inertial flow}

\begin{figure}
\hfill{}
\input{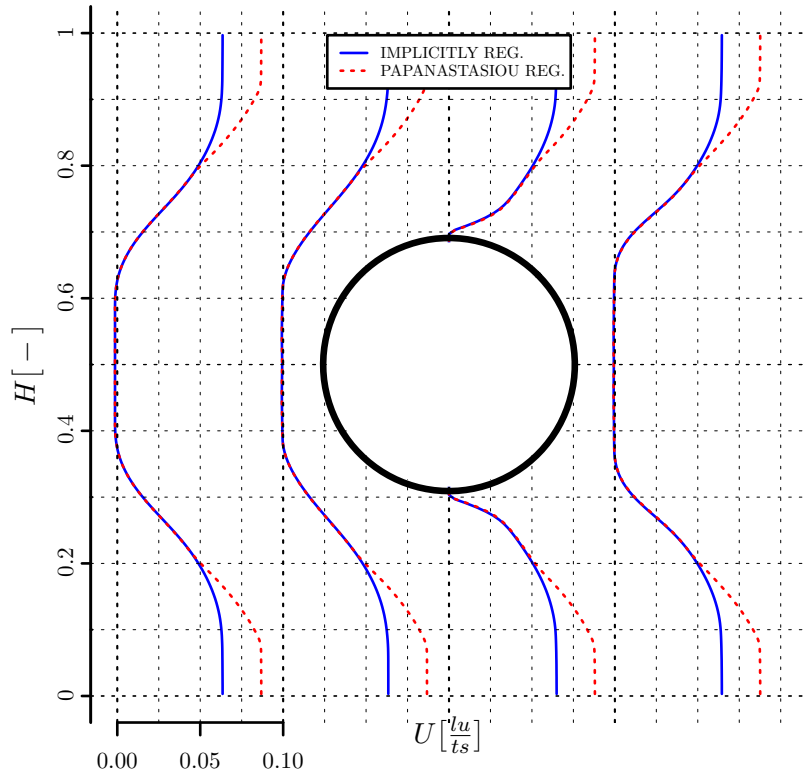}\hfill{}
\caption{Comparison of the horizontal velocity profiles for the PR and IR models for the third case listed in Table \ref{tab:ParticleInertial}, showing the excessive yielding and increased velocity of the PR model.}
\label{fig:Inertial160_comp}
\end{figure}

As a final comparison, both the PR and IR models were tested in the moderate Reynolds number flow regime, $Re\sim150-200$, for flows past a periodic array of cylinders. This investigation used the same configuration as that which was used in the creeping flow case, but here the driving force was increased in concert with the fluid yield stress. The plastic viscosity was set as $\nu_{p}=0.01$. Table \ref{tab:ParticleInertial} summarises the model details, results and comparison of the PR and IR models with the reference work \citep{Vikhansky2008_short_paper}. In the PR model it was found that the fluid yields excessively meaning the results for this lattice resolution were not valid owing to the Mach number constraint of the LBM (the maximal velocity obtained in the simulation exceeded $U=0.1$). Comparison of the PR and IR model for a larger lattice is discussed in Figure \ref{fig:Inertial160_comp}. A systematic discrepancy between the reference case was reported. Again, as in the creeping flow case, the flow is slower than in the reference case. The Strouhal number, $St=f_{v}D/U_{avg}$, was very similar for the first three cases whereas the flow in the fourth case does not destabilise at all. The flow instability was induced by a slight translation of the circular geometry resulting in an asymmetric staircase representation of the obstacle.

In this regime the PR and IR models show significant discrepancy, with the flow predicted by the PR model yielding considerably more than that predicted by the IR model. A detailed comparison of the velocity profiles for both models for one of the cases is shown in Figure \ref{fig:Inertial160_comp}. These simulations were performed on a two-times larger grid, $160\times160$, because the velocities attained by the PR models were too large for the LBM simulations on the smaller grid. Increasing the domain size resulted in lower velocities at the same value of plastic viscosity, $\nu_{p}=0.01$. The yield stress was set as $\sigma_{y}=1.1875\cdot10^{-5}$ and the body force intensity was $g=8.625\cdot10^{-7}$. The resulting similarity numbers for the IR model, $Re=201$, $Bn=2.422$, and $f=0.112$, are almost identical to those for the coarse resolution. Note that both cases are not exactly similar due to the staircase nature of the geometry representation, resulting in the particle being slightly smaller than in the $80\times80$ case. The similarity numbers for the PR model were found to be $Re=243$, $Bn=1.998$, and $f=0.076$.

\begin{table}
\begin{tabular}{|c|c|c|c|c|c|c|c|c|c|}
\hline 
 &  &  & \multicolumn{4}{c|}{Implicitly reg. (IR)} & \multicolumn{3}{c|}{\citet{Vikhansky2008_short_paper} (IR)}\tabularnewline
\hline 
\hline 
No.  & $\sigma_{y}$  & $g$  & $Re$  & $Bn$  & $f$  & $St$  & $Re$  & $Bn$  & $f$\tabularnewline
\hline 
1.  & $1.6\cdot10^{-5}$  & $4\cdot10^{-6}$  & $147$  & $1.11$  & $0.121$  & $0.399$  & $142$  & $1.13$  & $0.128$\tabularnewline
\hline 
2.  & $3.2\cdot10^{-5}$  & $5.4\cdot10^{-6}$  & $169$  & $1.93$  & $0.123$  & $0.377$  & $163$  & $1.97$  & $0.131$\tabularnewline
\hline 
3.  & $4.8\cdot10^{-5}$  & $6.9\cdot10^{-6}$  & $196$  & $2.51$  & $0.118$  & $0.390$  & $192$  & $2.49$  & $0.121$\tabularnewline
\hline 
4.  & $6.4\cdot10^{-5}$  & $7.9\cdot10^{-6}$  & $202$  & $3.14$  & $0.126$  & $-$  & $198$  & $3.23$  & $0.13$\tabularnewline
\hline 
\end{tabular}
\caption{Comparison of the similarity numbers in the case of viscoplastic inertial flow past a periodic array of cylinders in an $80\times80$ cell with a cylinder of $D=32lu$ and bounce-back boundary conditions.}
\label{tab:ParticleInertial}
\end{table}

The flow field in three out of four cases exhibited transient oscillations and the vortex shedding past the cylinder was observed. Unsteady flows were found that after a sufficiently long simulation time or due to a small asymmetry in the cylinder geometry. This is a typical phenomenon in Newtonian flow for Reynolds numbers starting from $Re\sim50$. The full von Karman vortex street is obviously not observed due to the periodicity of the model setup, which means that the vortices that are shed immediately hit another particle. However the vortex shedding frequency was measured by observing the sign of the vertical component of velocity behind the cylinder. The reference work did not address this issue and hence no comparison can be drawn.

\begin{figure}
\subfloat[$\sigma_{y}=1.6\cdot10^{-5},g=4\cdot10^{-6},Re=147$]{
\includegraphics[scale=0.30]{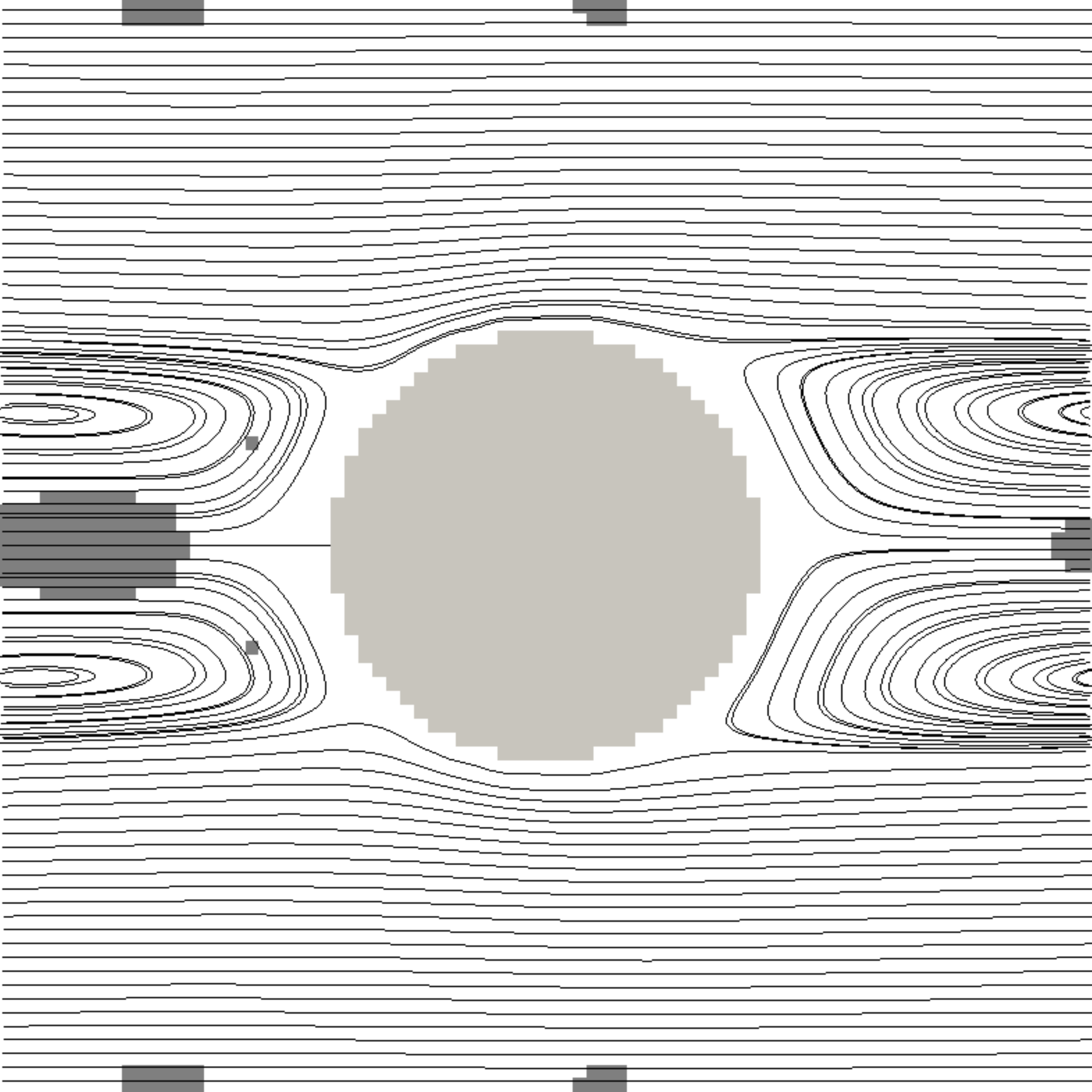}
}
\subfloat[$\sigma_{y}=3.2\cdot10^{-5},g=5.4\cdot10^{-6},Re=169$]{
\includegraphics[scale=0.30]{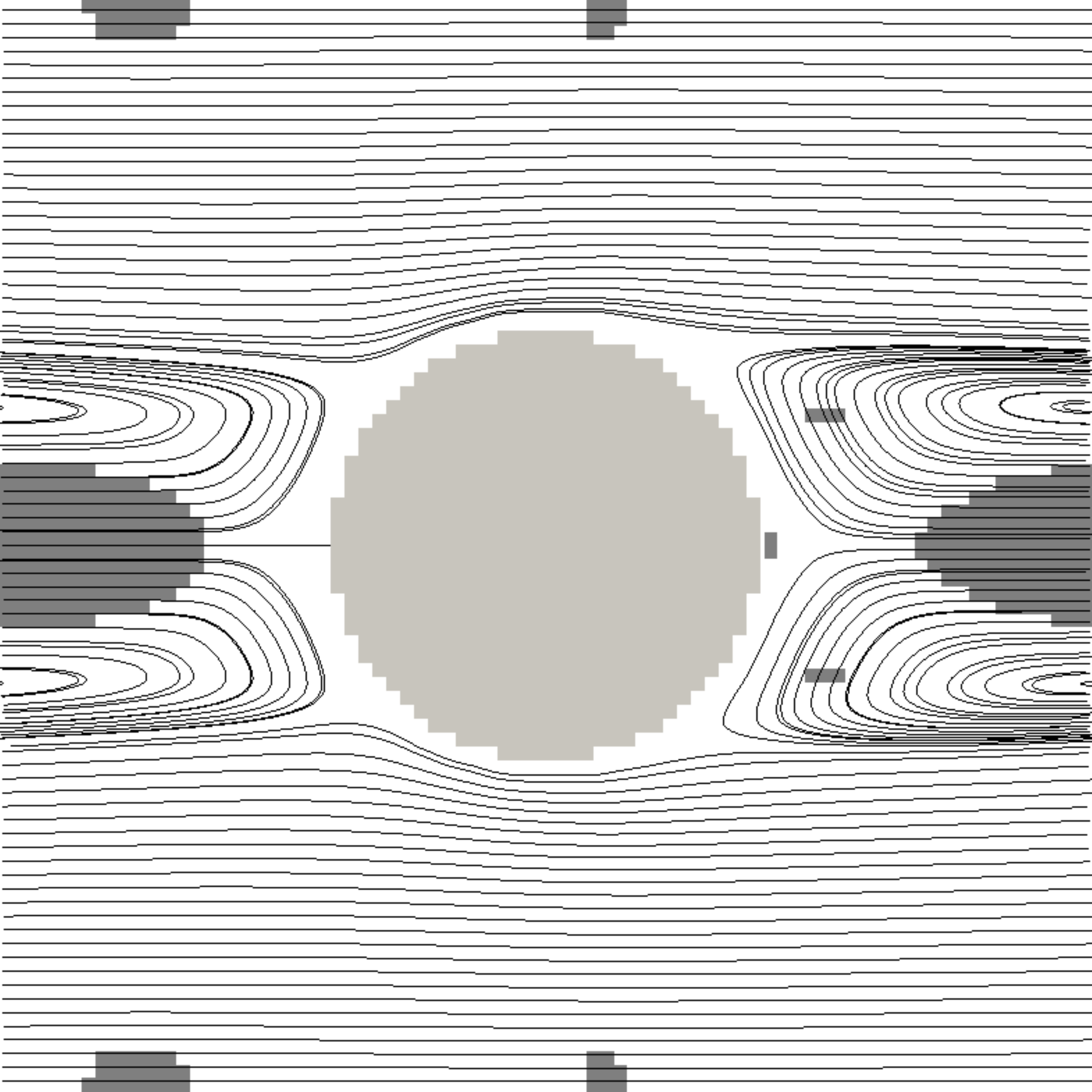}
}\\
\subfloat[$\sigma_{y}=4.8\cdot10^{-5},g=6.9\cdot10^{-6},Re=196$]{
\includegraphics[scale=0.30]{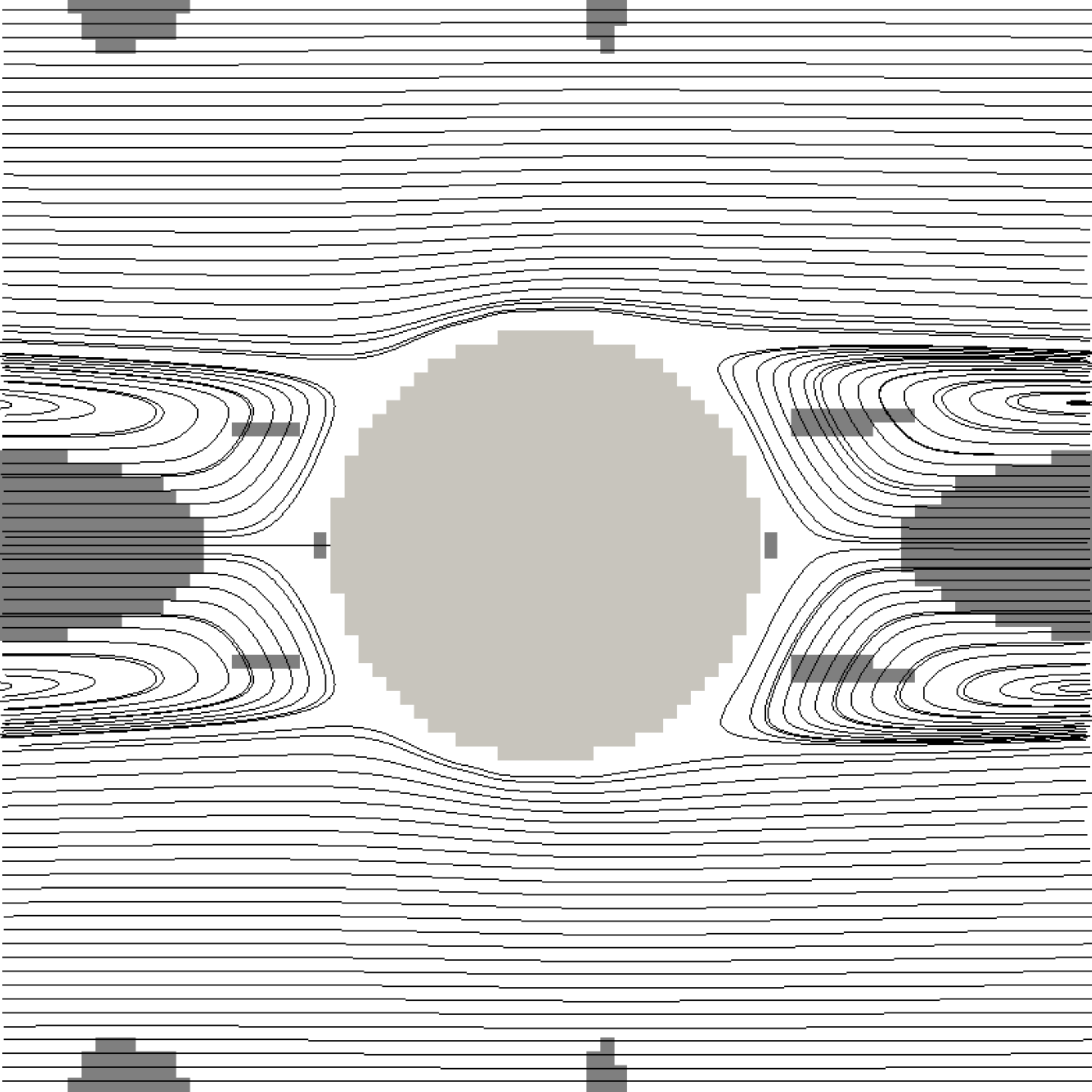}
}
\subfloat[$\sigma_{y}=6.4\cdot10^{-5},g=7.9\cdot10^{-6},Re=202$]{
\includegraphics[scale=0.30]{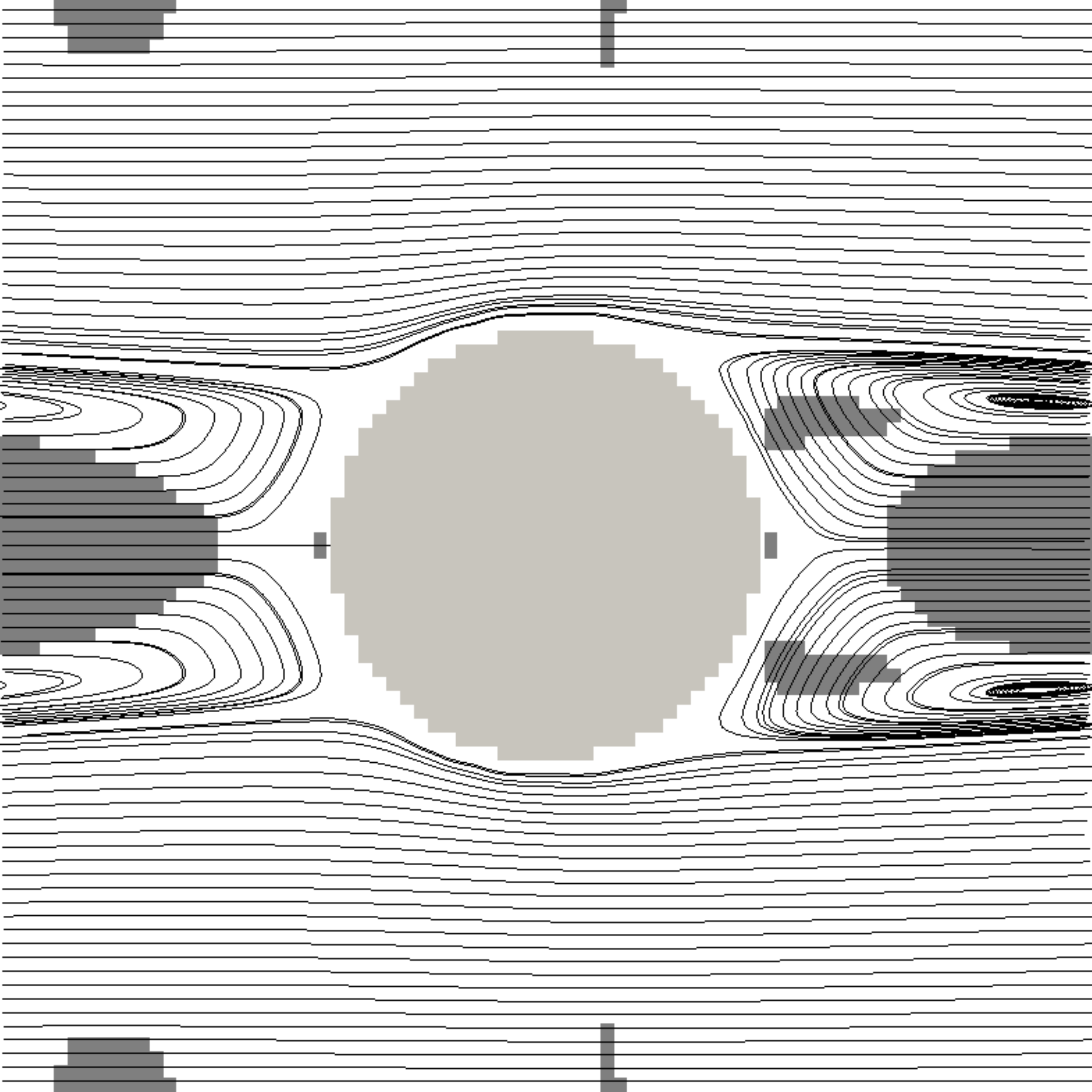}
}

\caption{Flow streamlines past a periodic array of cylinders in the inertial regime. The cylinder is marked with light grey and the unyielded zone with dark grey. All simulations were performed with the IR model on the $80\times80$ lattice.}
\label{fig:InertialSteadyFields}\end{figure}

Before the onset of vortex shedding, the flow fields remain in the
steady state for some time. The flow streamlines during the stable phase are shown in Figure \ref{fig:InertialSteadyFields}. In these it can be seen that a symmetric recirculation zone exists between the particles in all cases along with an unyielded zone which grows with increasing $\sigma_{y}$.

\section{Concluding remarks}
\label{sec:Conclusion}

In this paper, the performance of two LBM formulations for yield-stress (i.e. viscoplastic) fluids  has been investigated. The first, termed the PR model, uses Papanastasiou regularisation of the fluid rheology in conjunction with modification of the LBM relaxation rate based on the characteristic rate of strain. The second, termed the IR model, simultaneously solves for the fluid stress and underlying particle distribution functions at each grid point resulting in a type of locally implicit regularisation. In comparison of the two focus was given to issues, at least in the context of viscoplastic LBM, that have received little attention in the literature. This includes their spatial convergence, their susceptibility to spurious macroscopic velocities, and their behaviour in transient and inertial flows.

Initially, a comparison of the D3Q19 and D3Q27 lattice stencils using the standard Newtonian BGK and MRT models confirmed the presence of spurious velocity currents in the D3Q19 lattice stencil, as has been reported elsewhere. Consequently, the D3Q27 lattice was used for all subsequent simulations in this study. The influence of Burnett stresses, which are a consequence of the underlying kinetic construction of the LBM, was then investigated using the PR and IR models. For the PR model it was found that these non-hydrodynamic terms cause the Bingham apparent viscosity to deviate from the analytical solution at a regularisation parameter of approximately $10^9$ and higher. In the IR model, the Burnett stresses were found to induce small but spurious transverse fluid velocities which were largest in the vicinity of the yield surface. Despite these issues, it was found that the Burnett stresses did not significantly alter the macroscopic flow field and as such did not represent a barrier to either model in the simulation of viscoplastic fluids. Nevertheless, a generalised approach for the removal of these stresses from LBM simulations of Navier-Stokes flows remains desirable.

The spatial convergence of the PR and IR models was then demonstrated using creeping flows in circular and square ducts. For the IR model, this has not been previously demonstrated in such detail. Both analytical and semi-analytical solutions exist for the flow of Bingham fluids in circular and square ducts, respectively. However, in order to improve on the latter an alternative spectral solution was generated. In the case of Newtonian fluids, both the PR and IR models were found to demonstrate second order convergence in the square duct and first order convergence in the circular duct. The latter is attributed to known issues with the staircase representation of the boundary when using the bounce-back technique. The convergence behaviour of both models was found to be similar in the case of Bingham fluids. However, in the square duct it was found that the second order convergence of both models deteriorates as the lattice resolution increases. It is currently believed that this is a consequence of the imperfect accuracy of the spectral solution.

Following the convergence study, the transient performance of the PR and IR models in the square duct flows was interrogated. Here it was found that the IR model reproduced the flat velocity profile that is characteristic of Bingham fluids better than the PR model. Generally, the PR model was shown to suffer from considerable velocity fluctuations in the first $100$ time steps which at times resulted in part of the flow moving in the opposite direction to the driving body force. It is believed that this behaviour has implications for the use of the PR model in general transient and inertial flows.

Finally, the PR and IR models were compared in the case of flow past a periodic array of cylinders. Using the IR model and a Newtonian fluid it was shown that the calculated friction coefficient corresponded well with published data over a range of solid volume fractions. Using a Bingham fluid in the creeping flow regime and a range of yield stresses, it was found that the results of the PR model and IR model were almost identical. However, for the same models in the range of moderate Reynolds numbers, $Re\sim150$, it was found that the velocity profiles deviate between the PR and IR models. The PR model exhibits excessive yielding of the fluid which in turn results in higher fluid velocity, both of which represent a potential limitation of the PR model in inertial flows.

It has been demonstrated that the IR model performs better in transient and inertial flows of viscoplastic fluids than the PR model. Future work will further investigate the validation of the IR model in the case of moving boundaries and suspended particles. If successful, this will provide a foundation for the application of the IR model in the analysis of viscoplastic suspension rheology.

\section*{Nomenclature}

\subsection*{Latin letters}

$A_{i,j}$ - coefficients for the interpolating polynomials $\left(-\right)$

$Bn$ - Bingham number $\left(-\right)$

$\boldsymbol{c}_{i}$ - lattice velocity vector associated with $i$-th
distribution function $\left(-\right)$

$c_{i\alpha}$ - lattice velocity vector component $\left(-\right)$

$c_{s}$ - lattice speed of sound $\left(-\right)$

$D$ - diameter $\left(m,-\right)$

$d_{ij}$ - fluid strain rate tensor component $\left(\frac{1}{s},-\right)$

$\boldsymbol{D}$ - fluid strain rate tensor $\left(\frac{1}{s},-\right)$

$f$ - friction factor $\left(-\right)$

$f_{v}$ - frequency of vortex shedding $\left(\frac{1}{s}\right)$

$f_{i}$ - particle distribution function in the LBM simulations $\left(-\right)$

$\bar{f}_{i}$ - transformed particle distribution function in the
LBM simulations $\left(-\right)$

$\boldsymbol{f}$ - vector of all distribution functions in the LBM
simulations $\left(-\right)$

$\mathit{\boldsymbol{F}}$ - force acting on the fluid in the LBM
simulations $\left(-\right)$

$g$ - intensity of the body force $\left(\frac{m}{s^{2}},-\right)$

$H$ - channel height $\left(m,-\right)$

$Hg$ - Hagen number $\left(-\right)$

$\boldsymbol{I}$ - identity matrix $\left(-\right)$

$II_{D}$ - 2nd principal invariant of the strain rate tensor $\boldsymbol{D}$
$\left(\frac{1}{s},-\right)$

$II_{\sigma}$ - 2nd principal invariant of the stress tensor $\boldsymbol{\sigma}$
$\left(\frac{1}{s},-\right)$

$k$ - friction coefficient $\left(-\right)$

$L_{i}\left(\cdot\right)$ - Lagrange interpolating polynomial $\left(-\right)$

$m$ - regularisation parameter for the Papanastasiou-regularised
model $\left(-\right)$

$m_{i}$ - moment of particle distribution function $\left(-\right)$

$\mathit{\boldsymbol{m}}$ - vector of all moments of distribution
function $\left(-\right)$

$\mathit{\boldsymbol{M}}$ - MRT transformation matrix $\left(-\right)$

$Ma$ - Mach number $\left(-\right)$

$N$ - number of velocities in the lattice stencil, number of interpolation
nodes for the spectral solver $\left(-\right)$

$p$ - pressure in LBM $\left(-\right)$

$P_{n}^{\left(\alpha,\beta\right)}\left(\cdot\right)$ - Jacobi polynomial
of the $n-$th order  $\left(-\right)$

$q$ - density-specific pressure gradient/body force intensity driving the flow $\left(\frac{m}{s^{2}}\right)$

$Q$ - volumetric flow rate $\left(\frac{m^{2}}{s}\right)$

$R$ - radius $\left(m,-\right)$

$Re$ - Reynolds number $\left(-\right)$

$r_{y}$ - yield radius $\left(m,-\right)$

$s$ - contraction of the 2nd moment of the collision operator $\boldsymbol{s}$$\left(-\right)$

$\boldsymbol{s}$ - tensor of the 2nd moment of the collision operator $\left(-\right)$

$s_{i}$ - relaxation rates in the MRT collision operator $\left(-\right)$

$\boldsymbol{S}$ - diagonal matrix of LBM relaxation rates $\left(-\right)$

$St$ - Strouhal number $\left(-\right)$

$t$ - lattice time $\left(-\right)$

$\delta t$ - physical simulation time step $\left(s\right)$

$\Delta t$ - lattice time step $\left(-\right)$

$\boldsymbol{T}$ - 2nd order tensor of the non-equilibrium part of
the distribution function $\left(-\right)$

$\boldsymbol{u}$ - lattice fluid velocity vector $\left(-\right)$

$\boldsymbol{\Delta u}$ - lattice fluid velocity increase due to
external force $\left(-\right)$

$U$ - flow velocity $\left(-,\frac{m}{s}\right)$

$w_{i}$ - weights for the equilibrium distribution function $\left(-\right)$

$\boldsymbol{x}_{j}$ - vector of position of lattice node $\left(-\right)$

\subsection*{Greek letters}

$\epsilon$ - small parameter in the Chapman Enskog expansion $\left(-\right)$

$\varepsilon$ - relative error $\left(-\right)$

$\dot{\gamma}$ - characteristic rate of strain $\left(\frac{1}{s},-\right)$

$\Lambda$ - magic number $\left(-\right)$

$\mu$ - fluid dynamic viscosity $\left(\frac{kg}{m\cdot s}\right)$

$\mu_{app}$ - apparent fluid viscosity (dynamic) $\left(\frac{kg}{m\cdot s}\right)$

$\mu_{p}$ - fluid plastic viscosity (dynamic) $\left(\frac{kg}{m\cdot s}\right)$

$\nu$ - fluid kinematic viscosity $\left(\frac{m^{2}}{s},-\right)$

$\nu_{app}$ - apparent fluid viscosity (kinematic) $\left(\frac{m^{2}}{s},-\right)$

$\nu_{p}$ - fluid plastic viscosity (kinematic) $\left(\frac{m^{2}}{s},-\right)$

$\boldsymbol{\Pi}$ - second moment of the distribution function $\left(-\right)$

$\rho$ - lattice fluid density $\left(-\right)$

$\sigma$ - fluid stress tensor contraction $\left(Pa,-\right)$

$\boldsymbol{\sigma}$ - stress tensor $\left(Pa,-\right)$

$\sigma_{y}$ - fluid yield stress $\left(Pa,-\right)$

$\tau$ - lattice relaxation time $\left(-\right)$

$\omega$ - relaxation rate in the LBM simulation $\left(-\right)$

$\Omega\left(\mathbf{\mathit{\boldsymbol{f}}}\right)$ - LBM collision
operator $\left(-\right)$

\subsection*{Acknowledgments}

WR and CL wish to acknowledge The University of Queensland for the support of this research via UQ Early Career Researcher Grant (ECR090 2014003005). The comments of Dr. Tim Reis and Dr. Alexander Vikhansky during the preparation of this manuscript are also greatly appreciated.

\section*{References}

\bibliographystyle{myplainnat}
\bibliography{references}

\begin{thebibliography}{46}
\providecommand{\natexlab}[1]{#1}
\providecommand{\url}[1]{\texttt{#1}}
\expandafter\ifx\csname urlstyle\endcsname\relax
  \providecommand{\doi}[1]{doi: #1}\else
  \providecommand{\doi}{doi: \begingroup \urlstyle{rm}\Url}\fi

\bibitem[Barnes(1999)]{Barnes1999_panta_rei}
Barnes, H.~A.
\newblock The yield stress -- a review or "panta rei" -- everything flows?
\newblock \emph{Journal of Non-Newtonian Fluid Mechanics}, 81:\penalty0
  133--178, 1999.
\newblock \doi{10.1016/S0377-0257(98)00094-9}.
\newblock URL \url{http://dx.doi.org/10.1016/S0377-0257(98)00094-9}.

\bibitem[Bhatnagar et~al.(1954)Bhatnagar, Gross, and Krook]{BGK_original_paper}
Bhatnagar, P.~L., Gross, E.~P., and Krook, M.
\newblock A model for collision processes in gases. {I}. {S}mall amplitude
  processes in charged and neutral one-component systems.
\newblock \emph{Phys. Rev.}, 94:\penalty0 511--525, May 1954.
\newblock \doi{10.1103/PhysRev.94.511}.
\newblock URL \url{http://link.aps.org/doi/10.1103/PhysRev.94.511}.

\bibitem[Bingham(1916)]{Bingham_1916}
Bingham, E.~C.
\newblock An investigation of the laws of plastic flow.
\newblock \emph{Bulletin of the Bureau of Standards}, 13:\penalty0 309--353,
  1916.

\bibitem[Canuto et~al.(2006)Canuto, Hussaini, Quarteroni, and
  Zang]{CanutoQuarteroni2006Spectral}
Canuto, C., Hussaini, M., Quarteroni, A., and Zang, T.
\newblock \emph{Spectral Methods, Fundamentals in Single Domains}.
\newblock Springer, 2006.

\bibitem[Cercignani(1988)]{Cercignani1988}
Cercignani, C.
\newblock \emph{The {B}oltzmann equation and its applications}, volume~67 of
  \emph{Applied Mathematical Sciences}.
\newblock Springer New York, 1988.
\newblock ISBN 9780387966373.
\newblock \doi{10.1007/978-1-4612-1039-9}.

\bibitem[Chai et~al.(2011)Chai, Shi, Guo, and
  Rong]{Guo_et_all_2011_MRT_nonnewtonian}
Chai, Z., Shi, B., Guo, Z., and Rong, F.
\newblock Multiple-relaxation-time lattice {B}oltzmann model for generalized
  {N}ewtonian fluid flows.
\newblock \emph{Journal of Non-Newtonian Fluid Mechanics}, 166\penalty0
  (5-6):\penalty0 332 -- 342, 2011.
\newblock ISSN 0377-0257.
\newblock \doi{http://dx.doi.org/10.1016/j.jnnfm.2011.01.002}.
\newblock URL
  \url{http://www.sciencedirect.com/science/article/pii/S0377025711000073}.

\bibitem[Chen and Doolen(1998)]{Chen_Doolen_98_basicLBM}
Chen, S. and Doolen, G.~D.
\newblock {L}attice {B}oltzmann {m}ethod for fluid flows.
\newblock \emph{Annual Review of Fluid Mechanics}, 30\penalty0 (1):\penalty0
  329--364, 1998.
\newblock \doi{10.1146/annurev.fluid.30.1.329}.
\newblock URL \url{http://dx.doi.org/10.1146/annurev.fluid.30.1.329}.

\bibitem[Chen et~al.(2014)Chen, Zhang, Sun, and Jin]{China_2015_LBM_DEM}
Chen, S., Zhang, C., Sun, Q., and Jin, F.
\newblock Coupled {LBM}-{DEM} modeling of {Bingham} fluids with suspended
  particles.
\newblock In Soga, K., Kumar, K., Biscontin, G., and Kuo, M., editors,
  \emph{Geomechanics from {Micro} to {Macro}}, pages 479--484. CRC Press,
  August 2014.
\newblock ISBN 978-1-138-02707-7 978-1-315-73732-4.
\newblock URL \url{http://www.crcnetbase.com/doi/abs/10.1201/b17395-86}.

\bibitem[Chikatamarla et~al.(2010)Chikatamarla, Frouzakis, Karlin, Tomboulides,
  and Boulouchos]{ELB_turbulence_2010}
Chikatamarla, S.~S., Frouzakis, C.~E., Karlin, I.~V., Tomboulides, A.~G., and
  Boulouchos, K.~B.
\newblock Lattice {Boltzmann} method for direct numerical simulation of
  turbulent flows.
\newblock \emph{Journal of Fluid Mechanics}, 656:\penalty0 298--308, August
  2010.
\newblock ISSN 0022-1120, 1469-7645.
\newblock \doi{10.1017/S0022112010002740}.
\newblock URL
  \url{http://www.journals.cambridge.org/abstract\_S0022112010002740}.

\bibitem[Dellar(2003)]{Dellar_2003_MRT_convergence}
Dellar, P.~J.
\newblock Incompressible limits of lattice {B}oltzmann equations using multiple
  relaxation times.
\newblock \emph{J. Comput. Phys}, 2003.

\bibitem[Dellar(2013)]{Dellar2013_StrangSplitting}
Dellar, P.~J.
\newblock An interpretation and derivation of the lattice {B}oltzmann method
  using {S}trang splitting.
\newblock \emph{Computers \& Mathematics with Applications}, 65\penalty0
  (2):\penalty0 129 -- 141, 2013.
\newblock ISSN 0898-1221.
\newblock \doi{http://dx.doi.org/10.1016/j.camwa.2011.08.047}.
\newblock URL
  \url{http://www.sciencedirect.com/science/article/pii/S0898122111007206}.
\newblock Special Issue on Mesoscopic Methods in Engineering and Science
  (ICMMES-2010, Edmonton, Canada).

\bibitem[Dellar(2014)]{Dellar2014_Abstract2ndStress}
Dellar, P.~J.
\newblock Lattice {B}oltzmann formulation for linear viscoelastic fluids using
  an abstract second stress.
\newblock \emph{SIAM Journal on Scientific Computing}, 36\penalty0
  (6):\penalty0 A2507--A2532, 2014.
\newblock \doi{10.1137/130940372}.
\newblock URL \url{http://dx.doi.org/10.1137/130940372}.

\bibitem[Geier et~al.(2006)Geier, Greiner, and Korvink]{Cascaded_original_2006}
Geier, M., Greiner, A., and Korvink, J.~G.
\newblock Cascaded digital lattice {B}oltzmann automata for high {R}eynolds
  number flow.
\newblock \emph{Phys. Rev. E}, 73:\penalty0 066705, Jun 2006.
\newblock \doi{10.1103/PhysRevE.73.066705}.
\newblock URL \url{http://link.aps.org/doi/10.1103/PhysRevE.73.066705}.

\bibitem[Geier et~al.(2015)Geier, Sch\"onherr, Pasquali, and
  Krafczyk]{Geier2015_cumulant_LBM}
Geier, M., Sch\"onherr, M., Pasquali, A., and Krafczyk, M.
\newblock The cumulant lattice {B}oltzmann equation in three dimensions:
  {T}heory and validation.
\newblock \emph{Computers \& Mathematics with Applications}, 70\penalty0
  (4):\penalty0 507 -- 547, 2015.
\newblock ISSN 0898-1221.
\newblock \doi{http://dx.doi.org/10.1016/j.camwa.2015.05.001}.
\newblock URL
  \url{http://www.sciencedirect.com/science/article/pii/S0898122115002126}.

\bibitem[Geller et~al.(2013)Geller, Uphoff, and
  Krafczyk]{MRT_LESvs_Cascaded_jet_2013}
Geller, S., Uphoff, S., and Krafczyk, M.
\newblock Turbulent jet computations based on {MRT} and cascaded lattice
  {B}oltzmann models.
\newblock \emph{Computers \& Mathematics with Applications}, 65\penalty0
  (12):\penalty0 1956 -- 1966, 2013.
\newblock ISSN 0898-1221.
\newblock \doi{http://dx.doi.org/10.1016/j.camwa.2013.04.013}.
\newblock URL
  \url{http://www.sciencedirect.com/science/article/pii/S0898122113002198}.

\bibitem[Ginzburg and Steiner(2002)]{Ginzburg2002_Bingham_freesurface}
Ginzburg, I. and Steiner, K.
\newblock A free-surface lattice {B}oltzmann method for modelling the filling
  of expanding cavities by {B}ingham fluids.
\newblock \emph{Phil. Trans. R. Soc. Lond. A}, 360\penalty0 (1792):\penalty0
  453--466, 2002.
\newblock URL \url{http://dx.doi.org/10.1098/rsta.2001.0941}.

\bibitem[He and Luo(1997)]{LSL_1997_BEtoLBE}
He, X. and Luo, L.-S.
\newblock Theory of the lattice {B}oltzmann method: From the {B}oltzmann
  equation to the lattice {B}oltzmann equation.
\newblock \emph{Phys. Rev. E}, 56:\penalty0 6811--6817, Dec 1997.
\newblock \doi{10.1103/PhysRevE.56.6811}.
\newblock URL \url{http://link.aps.org/doi/10.1103/PhysRevE.56.6811}.

\bibitem[Higuera and Jim\'enez(1989)]{Higuera_Jimenez_1989}
Higuera, F.~J. and Jim\'enez, J.
\newblock Boltzmann approach to lattice gas simulations.
\newblock \emph{EPL (Europhysics Letters)}, 9\penalty0 (7):\penalty0 663, 1989.
\newblock URL \url{http://stacks.iop.org/0295-5075/9/i=7/a=009}.

\bibitem[Holmes et~al.(2011)Holmes, Williams, and
  Tilke]{Williams_Holmes_SPH_porous_2011}
Holmes, D.~W., Williams, J.~R., and Tilke, P.
\newblock Smooth particle hydrodynamics simulations of low {Reynolds} number
  flows through porous media.
\newblock \emph{Int. J. Numer. Anal. Meth. Geomech.}, 35\penalty0 (4):\penalty0
  419--437, 2011.
\newblock ISSN 1096-9853.
\newblock \doi{10.1002/nag.898}.
\newblock URL
  \url{http://onlinelibrary.wiley.com/doi/10.1002/nag.898/abstract}.

\bibitem[Huang et~al.(2011)Huang, Krafczyk, and Lu]{Krafczyk2011_forces_review}
Huang, H., Krafczyk, M., and Lu, X.
\newblock Forcing term in single-phase and {S}han-{C}hen-type multiphase
  lattice {B}oltzmann models.
\newblock \emph{Phys. Rev. E}, 84:\penalty0 046710, Oct 2011.
\newblock \doi{10.1103/PhysRevE.84.046710}.
\newblock URL \url{http://link.aps.org/doi/10.1103/PhysRevE.84.046710}.

\bibitem[{\swap{Humieres}{d'}} and Ginzburg(2009)]{MagickNumbers2009}
{\swap{Humieres}{d'}}, D. and Ginzburg, I.
\newblock Viscosity independent numerical errors for lattice {B}oltzmann
  models: {F}rom recurrence equations to "magic" collision numbers.
\newblock \emph{Computers \& Mathematics with Applications}, 58\penalty0
  (5):\penalty0 823 -- 840, 2009.
\newblock ISSN 0898-1221.
\newblock \doi{http://dx.doi.org/10.1016/j.camwa.2009.02.008}.
\newblock URL
  \url{http://www.sciencedirect.com/science/article/pii/S0898122109000893}.
\newblock Mesoscopic Methods in Engineering and Science.

\bibitem[{\swap{Humieres}{d'}} et~al.(2002){\swap{Humieres}{d'}}, Ginzburg,
  Krafczyk, Lallemand, and Luo]{MRT2002}
{\swap{Humieres}{d'}}, D., Ginzburg, I., Krafczyk, M., Lallemand, P., and Luo,
  L.-S.
\newblock Multiple-relaxation-time lattice {B}oltzmann models in three
  dimensions.
\newblock \emph{Phil. Trans. R. Soc. Lond. A}, 360:\penalty0 437--451, 2002.
\newblock URL
  \url{http://royalsocietypublishing.org/content/roypta/360/1792/437.full.pdf}.

\bibitem[{\swap{Humieres}{d'}} et~al.(2001){\swap{Humieres}{d'}}, Bouzidi, and
  Lallemand]{dHumieres_D3Q13_2001}
{\swap{Humieres}{d'}}, D., Bouzidi, M., and Lallemand, P.
\newblock Thirteen-velocity three-dimensional lattice {B}oltzmann model.
\newblock \emph{Phys. Rev. E}, 63:\penalty0 066702, May 2001.
\newblock \doi{10.1103/PhysRevE.63.066702}.
\newblock URL \url{http://link.aps.org/doi/10.1103/PhysRevE.63.066702}.

\bibitem[Kang and Hassan(2013)]{Texas_LBM_LES_spuriuos_2013}
Kang, S.~K. and Hassan, Y.~A.
\newblock The effect of lattice models within the lattice {B}oltzmann method in
  the simulation of wall-bounded turbulent flows.
\newblock \emph{Journal of Computational Physics}, 232\penalty0 (1):\penalty0
  100 -- 117, 2013.
\newblock ISSN 0021-9991.
\newblock \doi{http://dx.doi.org/10.1016/j.jcp.2012.07.023}.
\newblock URL
  \url{http://www.sciencedirect.com/science/article/pii/S0021999112003968}.

\bibitem[Karlin et~al.(2011)Karlin, Asinari, and Succi]{matrix_reloded_2011}
Karlin, I., Asinari, P., and Succi, S.
\newblock Matrix lattice {Boltzmann} reloaded.
\newblock \emph{Philosophical Transactions of the Royal Society A:
  Mathematical, Physical and Engineering Sciences}, 369\penalty0
  (1944):\penalty0 2202--2210, June 2011.
\newblock ISSN 1364-503X, 1471-2962.
\newblock \doi{10.1098/rsta.2011.0061}.
\newblock URL
  \url{http://rsta.royalsocietypublishing.org/cgi/doi/10.1098/rsta.2011.0061}.

\bibitem[Khirevich et~al.(2015)Khirevich, Ginzburg, and
  Tallarek]{Ginzburg_2015_spheres}
Khirevich, S., Ginzburg, I., and Tallarek, U.
\newblock Coarse- and fine-grid numerical behavior of {MRT}/{TRT}
  lattice-{B}oltzmann schemes in regular and random sphere packings.
\newblock \emph{Journal of Computational Physics}, 281:\penalty0 708 -- 742,
  2015.
\newblock ISSN 0021-9991.
\newblock \doi{http://dx.doi.org/10.1016/j.jcp.2014.10.038}.
\newblock URL
  \url{http://www.sciencedirect.com/science/article/pii/S0021999114007207}.

\bibitem[Kupershtokh et~al.(2009)Kupershtokh, Medvedev, and
  Karpov]{Kupershtokh2009}
Kupershtokh, A., Medvedev, D., and Karpov, D.
\newblock On equations of state in a lattice {B}oltzmann method.
\newblock \emph{Computers \& Mathematics with Applications}, 58\penalty0
  (5):\penalty0 965 -- 974, 2009.
\newblock ISSN 0898-1221.
\newblock \doi{http://dx.doi.org/10.1016/j.camwa.2009.02.024}.
\newblock URL
  \url{http://www.sciencedirect.com/science/article/pii/S0898122109001011}.

\bibitem[Kuwata and Suga(2015)]{Suga2015_stencil_errors}
Kuwata, Y. and Suga, K.
\newblock Anomaly of the lattice {B}oltzmann methods in three-dimensional
  cylindrical flows.
\newblock \emph{Journal of Computational Physics}, 280:\penalty0 563--569,
  2015.
\newblock ISSN 0021-9991.
\newblock \doi{http://dx.doi.org/10.1016/j.jcp.2014.10.002}.
\newblock URL
  \url{http://www.sciencedirect.com/science/article/pii/S0021999114006767}.

\bibitem[Leonardi et~al.(2014)Leonardi, Wittel, Mendoza, and
  Herrmann]{ALeonardi_2014_LBM_DEM_Binghamraey}
Leonardi, A., Wittel, F.~K., Mendoza, M., and Herrmann, H.~J.
\newblock Coupled {DEM-LBM} method for the free-surface simulation of
  heterogeneous suspensions.
\newblock \emph{Computational Particle Mechanics}, 1\penalty0 (1):\penalty0
  3--13, 2014.
\newblock ISSN 2196-4378.
\newblock \doi{10.1007/s40571-014-0001-z}.
\newblock URL \url{http://dx.doi.org/10.1007/s40571-014-0001-z}.

\bibitem[Leonardi et~al.(2011)Leonardi, Owen, and Feng]{Leonardi2011_power_law}
Leonardi, C.~R., Owen, D. R.~J., and Feng, Y.~T.
\newblock Numerical rheometry of bulk materials using a power law fluid and the
  lattice {B}oltzmann method.
\newblock \emph{Journal of Non-Newtonian Fluid Mechanics}, 166\penalty0
  (12-13):\penalty0 628 -- 638, 2011.
\newblock ISSN 0377-0257.
\newblock \doi{http://dx.doi.org/10.1016/j.jnnfm.2011.02.011}.
\newblock URL
  \url{http://www.sciencedirect.com/science/article/pii/S0377025711000772}.

\bibitem[Mosolov and
  Miasnikov(1966)]{Mosolov_Miasnikov_1966_limit_yield_surface_pipe_1}
Mosolov, P.~P. and Miasnikov, V.~P.
\newblock On stagnant flow regions of a viscous-plastic medium in pipes.
\newblock \emph{J. Mech. Appl. Math.}, 30:\penalty0 705--719, 1966.

\bibitem[Papanastasiou(1987)]{Papanastasiou1987}
Papanastasiou, T.~C.
\newblock Flows of materials with yield.
\newblock \emph{J. Rheol.}, 31\penalty0 (5):\penalty0 385--404, 1987.
\newblock URL \url{http://link.aip.org/link/?JOR/31/385/1}.

\bibitem[Reis and Dellar(2015)]{DellarReis_2016_BurnetStress}
Reis, T. and Dellar, P.
\newblock Burnett stress and the implementation of boundary conditions for
  lattice {B}oltzmann algorithms.
\newblock Paper presented at ICMMES 2015, 2015.

\bibitem[Sangani and Acrivos(1982)]{AcrivosSangani1982}
Sangani, A. and Acrivos, A.
\newblock Slow flow past periodic arrays of cylinders with application to heat
  transfer.
\newblock \emph{International Journal of Multiphase Flow}, 8\penalty0
  (3):\penalty0 193 -- 206, 1982.
\newblock ISSN 0301-9322.
\newblock \doi{http://dx.doi.org/10.1016/0301-9322(82)90029-5}.
\newblock URL
  \url{http://www.sciencedirect.com/science/article/pii/0301932282900295}.

\bibitem[Saramito and Rocquet(2001)]{Roquet_Saramito_2001_aFEM_pipe}
Saramito, P. and Rocquet, N.
\newblock An adaptive finite element method for viscoplastic fluid flows in
  pipes.
\newblock \emph{Comput. Methods Appl. Mech. Engrg.}, 190:\penalty0 5391--5412,
  2001.

\bibitem[Saramito and Rocquet(2003)]{Roquet_Saramito_2003_aFEM_cylinder}
Saramito, P. and Rocquet, N.
\newblock An adaptive finite element method for {B}ingham fluid flows around a
  cylinder.
\newblock \emph{Comput. Methods Appl. Mech. Engrg.}, 192:\penalty0 3317--3341,
  2003.
\newblock \doi{10.1016/S0045-7825(03)00262-7}.

\bibitem[Spelt et~al.(2005)Spelt, Yeow, Lawrence, and Selerland]{Spelt2005}
Spelt, P.~M., Yeow, A., Lawrence, C., and Selerland, T.
\newblock Creeping flows of {B}ingham fluids through arrays of aligned
  cylinders.
\newblock \emph{Journal of Non-Newtonian Fluid Mechanics}, 129\penalty0
  (2):\penalty0 66 -- 74, 2005.
\newblock ISSN 0377-0257.
\newblock \doi{http://dx.doi.org/10.1016/j.jnnfm.2005.05.007}.
\newblock URL
  \url{http://www.sciencedirect.com/science/article/pii/S0377025705001527}.

\bibitem[Stiebler et~al.(2011)Stiebler, Krafczyk, Freudiger, and
  Geier]{TUB_LES_sphere_2011}
Stiebler, M., Krafczyk, M., Freudiger, S., and Geier, M.
\newblock Lattice {B}oltzmann {L}arge {E}ddy {S}imulation of subcritical flows
  around a sphere on non-uniform grids.
\newblock \emph{Computers \& Mathematics with Applications}, 61\penalty0
  (12):\penalty0 3475 -- 3484, 2011.
\newblock ISSN 0898-1221.
\newblock \doi{http://dx.doi.org/10.1016/j.camwa.2011.03.063}.
\newblock URL
  \url{http://www.sciencedirect.com/science/article/pii/S0898122111002264}.
\newblock Mesoscopic Methods for Engineering and Science - Proceedings of
  ICMMES 09 Mesoscopic Methods for Engineering and Science.

\bibitem[Suga et~al.(2015)Suga, Kuwata, Takashima, and
  Chikasue]{Suga2015_MRT_LES}
Suga, K., Kuwata, Y., Takashima, K., and Chikasue, R.
\newblock A {D}3{Q}27 multiple-relaxation-time lattice {B}oltzmann method for
  turbulent flows.
\newblock \emph{Computers \& Mathematics with Applications}, 69\penalty0
  (6):\penalty0 518 -- 529, 2015.
\newblock ISSN 0898-1221.
\newblock \doi{http://dx.doi.org/10.1016/j.camwa.2015.01.010}.
\newblock URL
  \url{http://www.sciencedirect.com/science/article/pii/S0898122115000346}.

\bibitem[Svec et~al.(2012)Svec, Skocek, Stang, Geiker, and
  Roussel]{Svec2012Denmark}
Svec, O., Skocek, J., Stang, H., Geiker, M.~R., and Roussel, N.
\newblock Free surface flow of a suspension of rigid particles in a
  non-{N}ewtonian fluid: A lattice {B}oltzmann approach.
\newblock \emph{Journal of Non-Newtonian Fluid Mechanics}, 179-180:\penalty0 32
  -- 42, 2012.
\newblock ISSN 0377-0257.
\newblock \doi{http://dx.doi.org/10.1016/j.jnnfm.2012.05.005}.
\newblock URL
  \url{http://www.sciencedirect.com/science/article/pii/S0377025712001061}.

\bibitem[Taylor and Wilson(1997)]{TaylorWilsonBicr112Reg}
Taylor, A. and Wilson, D.
\newblock Conduit flow of an incompressible, yield-stress fluid.
\newblock \emph{J. Rheol.}, 41\penalty0 (93):\penalty0 93--101, 1997.

\bibitem[Vikhansky(2008)]{Vikhansky2008_short_paper}
Vikhansky, A.
\newblock Lattice-{B}oltzmann method for yield-stress liquids.
\newblock \emph{Journal of Non-Newtonian Fluid Mechanics}, 155\penalty0
  (3):\penalty0 95 -- 100, 2008.
\newblock ISSN 0377-0257.
\newblock \doi{http://dx.doi.org/10.1016/j.jnnfm.2007.09.001}.
\newblock URL
  \url{http://www.sciencedirect.com/science/article/pii/S0377025707001978}.

\bibitem[Vikhansky(2012)]{Vikhansky_2011_Canada}
Vikhansky, A.
\newblock Construction of lattice-{B}oltzmann schemes for non-{N}ewtonian and
  two-phase flows.
\newblock \emph{The Canadian Journal of Chemical Engineering}, 90\penalty0
  (5):\penalty0 1081--1091, 2012.
\newblock ISSN 1939-019X.
\newblock \doi{10.1002/cjce.21664}.
\newblock URL \url{http://dx.doi.org/10.1002/cjce.21664}.

\bibitem[Wang and Ho(2008)]{China_2008_rheology_in_FEQ}
Wang, C.-H. and Ho, J.-R.
\newblock Lattice {B}oltzmann modeling of {B}ingham plastics.
\newblock \emph{Physica A: Statistical Mechanics and its Applications},
  387\penalty0 (19-20):\penalty0 4740 -- 4748, 2008.
\newblock ISSN 0378-4371.
\newblock \doi{http://dx.doi.org/10.1016/j.physa.2008.04.008}.
\newblock URL
  \url{http://www.sciencedirect.com/science/article/pii/S0378437108003683}.

\bibitem[Wang et~al.(2015)Wang, Mi, Meng, and Guo]{Guo2015_nonNewtInFeq}
Wang, L., Mi, J., Meng, X., and Guo, Z.
\newblock A localized mass-conserving lattice {B}oltzmann approach for
  non-{N}ewtonian fluid flows.
\newblock \emph{Communications in Computational Physics}, 17:\penalty0
  908--924, 4 2015.
\newblock ISSN 1991-7120.
\newblock \doi{10.4208/cicp.2014.m303}.
\newblock URL \url{http://journals.cambridge.org/article\_S1815240615000249}.

\bibitem[Zhu et~al.(1999)Zhu, Fox, and Morris]{Zhu_1999}
Zhu, Y., Fox, P.~J., and Morris, J.~P.
\newblock A pore-scale numerical model for flow through porous media.
\newblock \emph{International Journal for Numerical and Analytical Methods in
  Geomechanics}, 23\penalty0 (9):\penalty0 881--904, August 1999.
\newblock
  \doi{10.1002/(SICI)1096-9853(19990810)23:9<881::AID-NAG996>3.0.CO;2-K}.
\newblock URL
  \url{http://doi.wiley.com/10.1002/\%28SICI\%291096-9853\%2819990810\%2923\%3A9\%3C881\%3A\%3AAID-NAG996\%3E3.0.CO\%3B2-K}.

\end{thebibliography}

\end{document}